\documentclass[prodmode,acmtochi]{acmsmall}

\acmVolume{1}
\acmNumber{1}
\acmArticle{1}
\articleSeq{1}
\acmYear{2018}
\acmMonth{1}

\setcopyright{rightsretained}

\usepackage[ruled]{algorithm2e}
\SetAlFnt{\algofont}
\SetAlCapFnt{\algofont}
\SetAlCapNameFnt{\algofont}
\SetAlCapHSkip{0pt}
\IncMargin{-\parindent}
\renewcommand{\algorithmcfname}{ALGORITHM}

\usepackage{graphics}
\usepackage{multirow}
\usepackage{array}
\usepackage{hhline}
\graphicspath{{figs/}}

\title{Tools for online tutorials: comparing capture devices, tutorial representations, and access devices}
\author{
  Scott Carter, Pernilla Qvarfordt, Matthew Cooper, Aki Komori
    \affil{FX Palo Alto Laboratory, Palo Alto, CA USA}
  Ville M\"{a}kel\"{a}
    \affil{TAUCHI, School of Information Sciences, University of Tampere, Finland } 
}

\begin{abstract}
Tutorials are one of the most fundamental means of conveying knowledge. Ideally when the task involves physical or digital objects, tutorials not only {\em describe} each step with text or via audio narration but {\em show} it as well using photos or animation. In most cases, online tutorial authors capture media from handheld mobile devices to compose these documents, but increasingly they use wearable devices as well. In this work, we explore the full life-cycle of online tutorial creation and viewing using head-mounted capture and displays.
We developed a media-capture tool for Google Glass that requires minimal attention to the capture device and instead allows the author to focus on creating the tutorial's content rather than its capture. The capture tool is coupled with web-based authoring tools for creating annotatable videos and multimedia documents. 
In a study comparing standalone (camera on tripod) versus wearable capture (Google Glass) as well as two types of multimedia representation for authoring tutorials (video-based or document-based), we show that tutorial authors have a preference for wearable capture devices, especially when recording activities involving larger objects in non-desktop environments. Authors preferred document-based multimedia tutorials because they are more straightforward to compose and the step-based structure translates more directly to explaining a procedure.

In addition, we explored using head-mounted displays (Google Glass) for accessing tutorials in comparison to lightweight computing devices such as tablets. Our study included tutorials recorded with the same capture methods as in our access study. We found that although authors preferred head-mounted capture, tutorial consumers preferred video recorded by a camera on tripod that provides a more stable image of the workspace. Head-mounted displays are good for glanceable information, however video demands more attention and our participants made more errors using Glass than when using a tablet, which was easier to ignore. Our findings point out several design implications for online tutorial authoring and access methods.

\end{abstract}
\terms{Human Factors}

\begin{document}

\begin{bottomstuff}
Author's address: S. Carter, P. Qvarfordt, A. Komori, M. Cooper , FX Palo Alto Laboratory, 3174 Porter Dr., Palo Alto, CA 94304; email:
\{carter,pernilla,komori,cooper\}@fxpal.com; Ville M\"{a}kel\"{a}, TAUCHI, School of Information Sciences, University of Tampere, Finland; email:ville.mi.makela@sis.uta.fi 
\end{bottomstuff}

\maketitle

\section{Introduction}

Online how-tos are popular for a diverse range of topics, including coding, cooking, and maintenance. The lifecycle of an online tutorial involves an author capturing the steps necessary to complete a task, and then assembling, editing, and annotating a curated set of captured material into a multimedia representation. Finally, an end user will access the tutorial and apply the instructions to aid in completing a similar task. However, there are considerable variations in each step. The tutorial author can choose among many different capture and authoring methods as well as representation approaches. Additionally, the end user can potentially access the tutorial in a variety of ways, some of which may not align well with the methods the author chose to capture and represent the task. 

Furthermore, relying on a single media type is rarely the best way to convey expository content. 
For certain tasks, video has been shown to be particularly helpful beyond static graphics \cite{Grossman,Pongnumkul}. This is intuitive since some tasks involve gradual progressions that can be difficult to capture in static photos (e.g., fluffing egg whites). Other tasks can require multimedia feedback (e.g., playing musical instruments). 

Video can also 
help coordinate a set of steps into a cohesive sequence. 
For example, the act of kicking a football can be 
depicted by a series of static shots: lining up the foot, striking the ball at a particular spot, following through, etc.  Without seeing these individual elements combined in a swift strike, it can be difficult to verify the correctness of the composite end result. Furthermore, using video does not preclude integrating static content -- many video editing tools support the integration of static photos that can be ``played'' for some period of time within the video. Semi-automated systems can also help condense expository video into more consumable clips \cite{Chi13}. However, past work has shown that video is not always the best presentation format for all learning tasks \cite{Palmiter}. In some cases the best approach to show how to accomplish a task involves combining text with static graphics \cite{Harrison}.  



In this work we investigate different methods of multimedia tutorial capture, representation, and access.  For tutorial authors we compare two different multimedia capture approaches: mobile devices and head-mounted devices. Formative work we conducted with tutorial authors showed that common mobile devices are the most common capture approach for these types of videos. Our formative work also indicated that many tutorial authors had difficulty recording complicated procedures with standard mounted-cameras, suggesting a role for head-mounted capture. This approach stands to benefit end users as well since past work has suggested that first-person video instruction can improve performance on assembly \cite{Kraut} and learning \cite{Lindgren} tasks. Our toolset includes a head-mounted capture application to realize these advantages that was the focus of earlier work~\cite{carter2015creating}.     

Creating effective how-tos also requires the means to mix different types of media within the authoring process.  Our toolset includes a variety of authoring and multimedia annotation capabilities.  We focus on supporting two authoring metaphors: temporal (video-based) and spatial (document-based).  We support both approaches to composing how-tos and also allow for mixing them together. Our tools also facilitate flexible access to how-to content with the use of annotations in the form of bookmarks that optionally include or link to supplementary multimedia information. Our tools can also use tags to group bookmarks and their associated content to aid in both retrieval and the authoring process.  The authoring process endows the resulting content with structure that is exploited in a head-mounted device access tool which we evaluate in this paper.


In the next section, we describe previous work and needfinding results that drove the development of this toolset. We then show how the tools we built help authors create tutorials and describe the relative strengths and weaknesses of different capture and authoring approaches for different types of tasks.

\section{Tutorial capture and authoring related work} 

To understand how to support authors creating expository multimedia content, we investigated past work focusing on the role of media in learning and knowledge transfer. We also investigated first-hand, participatory observations of the use of off-the-shelf capture and access tools.

\subsection{Multimedia Instructions}

Tutorials can enhance different kinds of learning, learning by example or learning by principle. Eiriksdottir and Catrambone conducted a review of instructions for procedural tasks  \cite{Eiriksdottir}. They suggest that specific procedural instructions grounded with realistic examples and limited reference to more general principles 
produce better primary task performance but poor learning and transfer to other tasks. 
In many cases, users turn to a how-to or tutorial video to obtain specific information without particularly needing or wanting to learn about more general principles -- for example, when fixing their printer. Thus it is critical that tutorial structure supports initial performance, implying a focus on step-by-step instructions. Other work has shown that higher quality examples enhance task performance \cite{Pirolli} and that learning can improve with the incorporation of video-based examples in particular \cite{Moreno}. Coupled with Clark's and Mayer's finding that multimedia is especially useful for ``learners who have low knowledge of a domain'' \cite{Clark}, this suggests that tools for creating tutorial and how-to video should support links to concrete examples and complementary multimedia materials. 

However, it is important that the tutorial presentation tool aid users in acquiring knowledge transferable to other tasks and domains. For this, presentation tools should support users actively navigating \cite{Schwan} as well as annotating and editing to form their own interpretations of video content \cite{Zahn} \cite{Barthel}. Zhang {\em et al.} \cite{Zhang} found that interactive video in particular ``achieved significantly better learning performance'' than linear video because 1) content can be repeated; 2) the interface enables random access, which ``is expected to increase learner engagement'' and allows the user to control the pace of learning; and 3) it can increase learner attentiveness. 

Overall, past work suggests that interactive video complemented with rich multimedia materials and specific examples can help users complete short-term tasks while potentially developing transferable knowledge. 

\subsection{Tutorial Creation} 



It can be difficult to understand multimedia tutorial and how-to guide authoring practices, since authors and viewers vary from non-expert videographers to more advanced content creators. Torrey {\em et al.} \cite{Torrey07} interviewed authors of how-to videos to explore how their guides are produced and distributed. They found that authors utilize an array of tools and broadcast methods to construct, iterate, and diffuse their guides. Specifically, they found that users combined video, annotated photos, text, diagrams, and other media to communicate their work. These findings suggest that tools to support how-to creation and search should incorporate multimedia and mark-up.  
Torrey {\em et al.} \cite{Torrey07,Torrey09} explored how electronics and computer hobbyists create how-to content. 

\cite{Carter}, Carter {\em et al.} interviewed nine participants about their how-to creation practices and were able to observe two participants as they jointly created a how-to guide.  
They found that authors need to be able to choose a capture device that supports the context of the activity -- what is being documented and where -- so that media capture can remain as unobtrusive as possible.
The spatial context of an activity introduces key constraints into content capture.
Table-top tutorials tend to focus on the construction of smaller objects and involve fewer camera-angle changes and longer shots. On the other hand, how-tos concerning other tasks (e.g., automotive, home repair, etc.) necessarily require a wider variety of camera angles and shorter duration shots.

They also reported that authors had difficulty creating a narrative using video exclusively. Many participants wanted to import and link to external content as well as bookmark and annotate important sections of their video. Participants also wanted to link dynamically to other examples or versions of the object or activity they were documenting.


In summary, Torrey {\em et al.} found that how-to ``sharing occurs within and across a collection of communication tools without any centralized control'' \cite{Torrey07} and that people tend to find information by browsing as much as by more directed search \cite{Torrey09}. Carter {\em et al.} previously found that tools for the capture, creation, and access of how-to guides were similarly decentralized \cite{Carter}. 

 

\section{ShowHow overview}
In response, we built ShowHow, a suite of capture and authoring tools to support building tutorials. 
%
%
ShowHow users can capture content with either the ShowHow mobile app or a third-party mobile client of their choosing, and use ShowHow's HTML5-based drag-and-drop authoring tools to assemble captured media into a tutorial. As noted above, we found that handheld capture devices are too intrusive for many applications as they require users to focus on the device rather than the subject of the how-to they are authoring. For this reason we choose to explore head-mounted capture of and access to tutorials.

\subsection{Head-mounted capture}

\begin{figure*}[tbh]
\begin{tabular}{c}
\includegraphics[scale=.135]{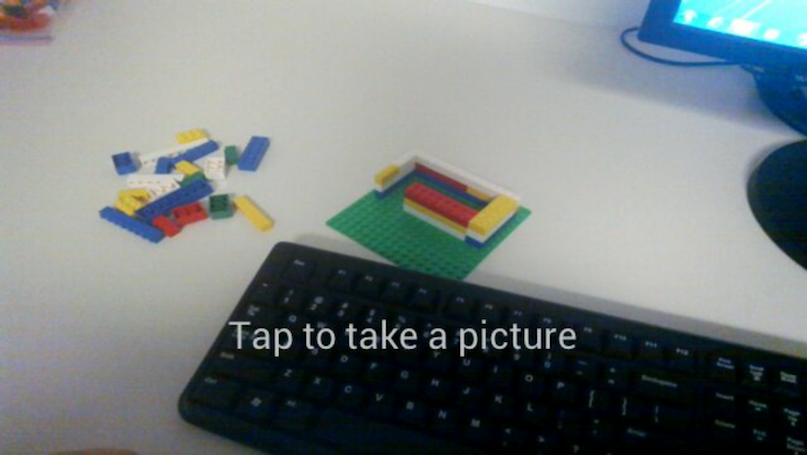} \\
a \\
\includegraphics[scale=.27]{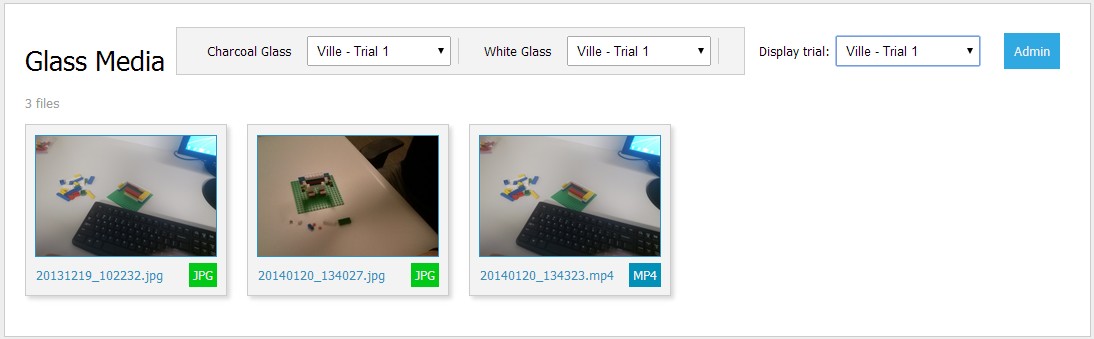} \\
b \\
\begin{tabular}{cc}
\includegraphics[scale=.189]{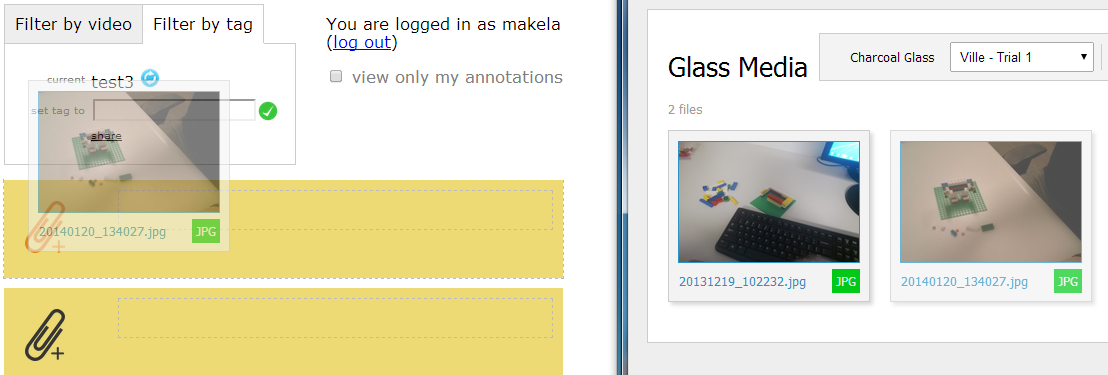} &
\includegraphics[scale=.189]{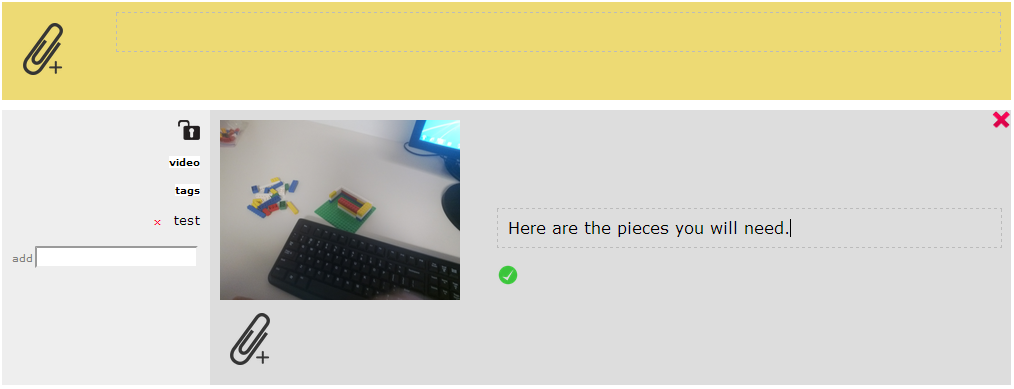} \\
c & d
\end{tabular}
\end{tabular}
\caption{Capturing media with the ShowHow Glass application. Authors snap a photo or record a video with the Glass application (a). Media is automatically synchronized to the ShowHow web server and client (b). From there, authors can drag media into the authoring tools (c). Next, the author can add text annotations describing the captured content (d).}
\label{fig:shGlass}
\end{figure*}

We built a new head-mounted capture application for Google Glass, which couples a head-mounted camera and display. Crucially, the Glass display sits on the periphery of a user's visual field, allowing them to focus naturally on 
completing the task to be documented. With a quick glance at the peripheral display, the user can confirm the state of the recording while their hands remain free for the activity of interest. A head-mounted camera is also likely easier for the author to manipulate than a mobile device.
Our aim was to provide a capture tool that would require minimal interaction and allow how-to authors to focus on their message rather than the tool.

Because of limitations of Glass at the time of our experiments, we developed our own capture application for Google Glass enabling users to take pictures and record videos without restrictions and that automatically uploads content to the ShowHow server (Figure \ref{fig:shGlass}).
The Glass application is built on Android platform version 4.0.3. It uses the Glass Development Kit library's gesture detector to recognize different gestures on the Glass touchpad. A separate component handles uploads to the server.

The application consists of three views. From the main menu, users select whether they want to take pictures or record videos. To navigate, the user swipes forward or backwards on the touchpad attached to Glass' frame and taps to select. Users return to the main menu by swiping down on the touchpad. In picture view a single tap takes a picture, while in video view the first tap starts and the second tap stops the recording (Step A in Figure \ref{fig:shGlass}). Glass' display shows a viewfinder while in capture mode, making the recording glanceable. Media is uploaded automatically, and the upload progress is updated in the top-right corner of the screen of Google Glass. Multiple files can be uploaded simultaneously. The Glass application uploads files asynchronously and they appear in the web application as they become available. After media are uploaded, they are visible in ShowHow (Step B in Figure \ref{fig:shGlass}) in the order they became available. Now media can be dragged-and-dropped into a ShowHow tool for authoring (Step C \& D in Figure \ref{fig:shGlass}). In this way, authors can integrate captures from Glass into either the video-based or document-based tutorials described below.

\subsection{Authoring how-tos with ShowHow}

ShowHow supports two separate methods for authoring: \emph{video-based} and \emph{document-based}. How-to content is commonly structured according to the {\em steps} that comprise the documented activity.  Both authoring approaches allow users to combine different types of media to craft 
tutorials that optionally include step boundaries, step sequence, step importance, and references to other activities beyond the immediate scope of the current how-to. 
The key difference is that the video-based approach fundamentally relies on a {\em temporal} organizational metaphor wherein the system {\em plays through} the video and associated bookmarks, while the document-based approach relies on a {\em spatial} organizational metaphor wherein the user {\em scrolls through} content (similar to \cite{Chi}). 

\subsubsection{Making a video-based how-to with ShowHow}

 \begin{figure}[tbh]
 \begin{center}
 \begin{tabular}{cc} 
  \includegraphics[width=2.5in]{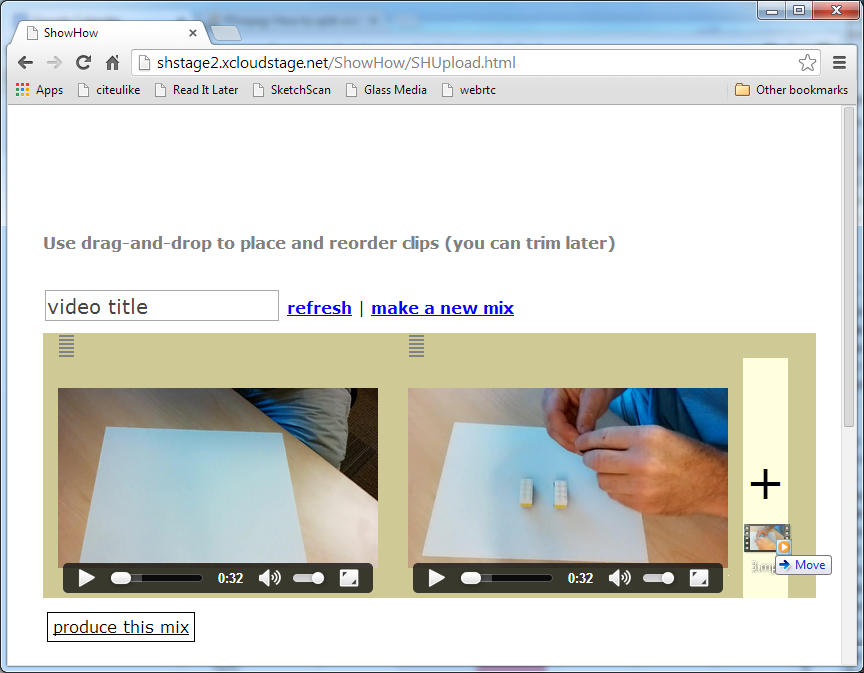} & \includegraphics[width=2.5in]{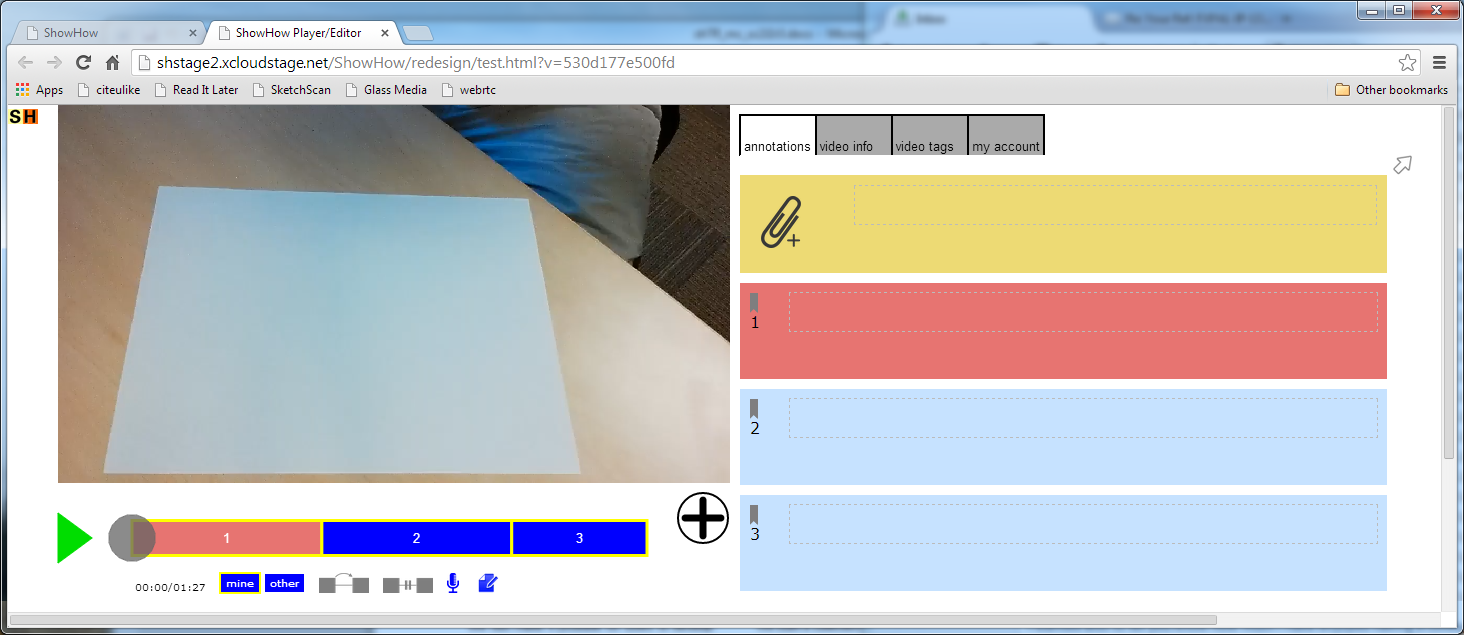} \\
 (a) & (b)  \\
 \end{tabular}
 \end{center}
 \caption{Creation process with ShowHow for video-based tutorials. (a) An author drags clips into the create/mix view. Clips are uploaded to the server, which begins transcoding them asynchronously. Authors can rearrange the order of clips via drag-and-drop. Once satisfied with the clip collection, the author can produce the mix. (b) Once it is ready, any user can view the video tutorial. Note that the system created three bookmarks, one for each input clip.}
 \label{fig:videoEdit}
 \end{figure}

\begin{figure*}[tbh]
\begin{center}
\begin{tabular}{ccc} 
 \includegraphics[width=1.53in]{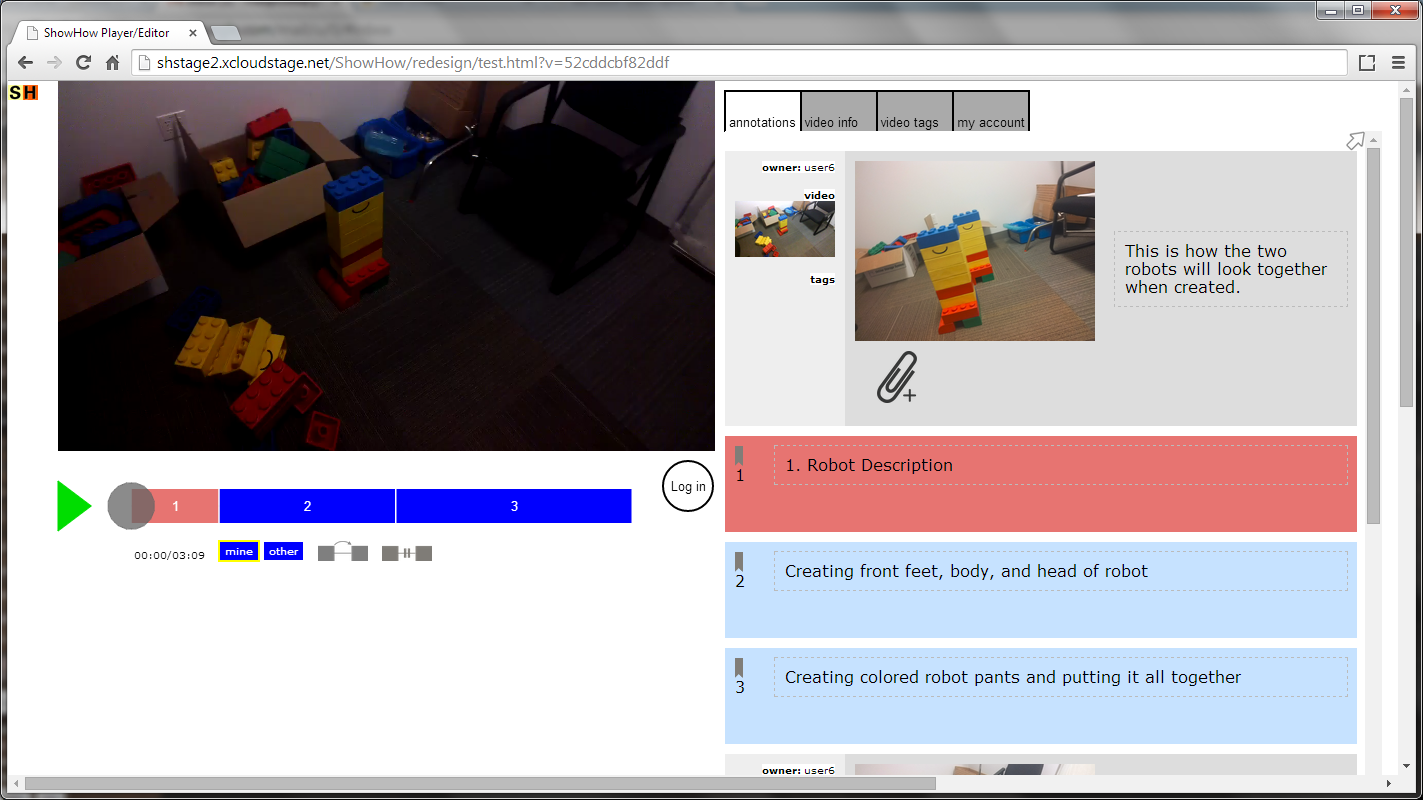} &
 \includegraphics[width=1.53in]{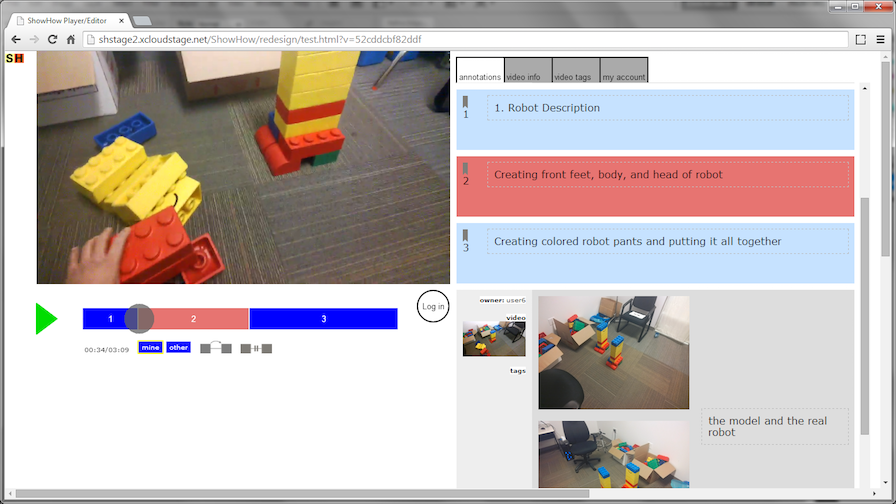} &
 \includegraphics[width=1.53in]{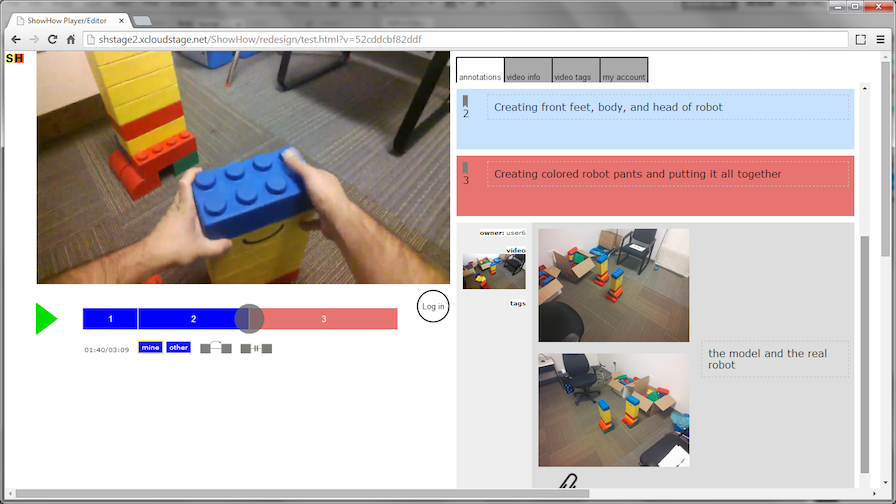} \\
\end{tabular}
\end{center}
\caption{A video-based tutorial created by a participant in the study. This tutorial includes three steps with a bookmark for each step as well as annotations at the beginning and end with photos showing the full figure.}
\label{fig:videoExample}
\end{figure*}

The process of creating a video-based how-to involves 1) uploading raw content; 2) editing content; and 3) adding bookmarks and multimedia annotations to augment the how-to. ShowHow's video-based user interface consists of two views built in HTML-5. 
In the create/mix view, authors can drag-and-drop video clips onto the tool, drag to reorder the clips, and play each clip individually (see Figure \ref{fig:videoEdit} a). This approach allows authors to combine clips rapidly. 
When the author is satisfied with the order and edits of individual clips she clicks a button to create a composite clip. 

Once processing is complete, ShowHow changes automatically to the video-player view where users can view the video and begin adding markup (Figure \ref{fig:videoEdit}b). The video-player view has a video player and a video timeline to the left, and a markup panel to the right. To start off the mark-up, ShowHow automatically creates one bookmark for each individual video clip. The bookmarks appear as blue rectangles in Figure \ref{fig:videoEdit}b, the currently selected video clip and bookmark is salmon pink. 


Besides bookmarks, ShowHow also supports another basic type of video markup - annotations. Both can add supporting text to the tutorial. The main differences are that 1) bookmarks are tied to the composite video clip, having a start time and, optionally, an end time to define a video segment, and 2) that annotations can have an arbitrary number of media and tags. Because bookmarks indicate a time span in a video they are represented both on the video timeline as well as in the markup panel. 

\begin{figure}[tb]
\centering
\begin{tabular}{cc} 
 \includegraphics[width=2.54in]{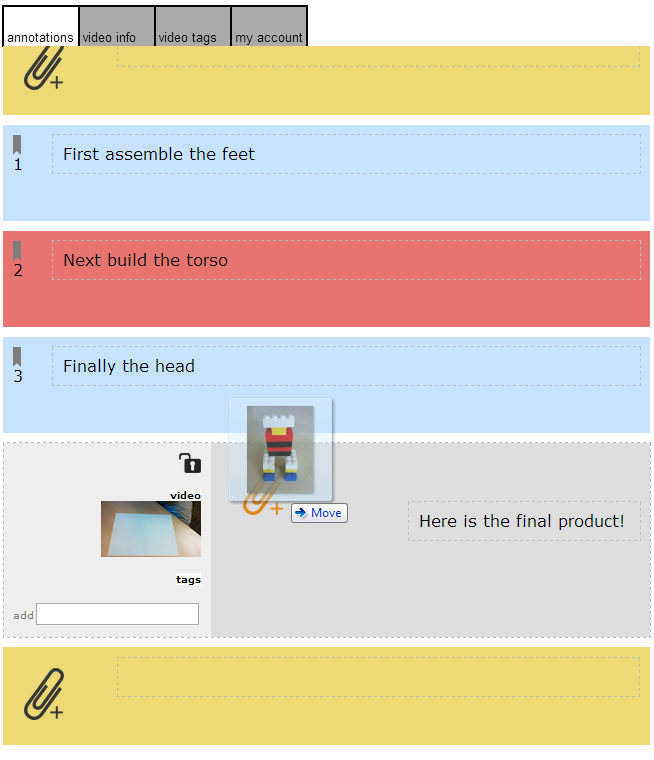} &
 \includegraphics[width=2.54in]{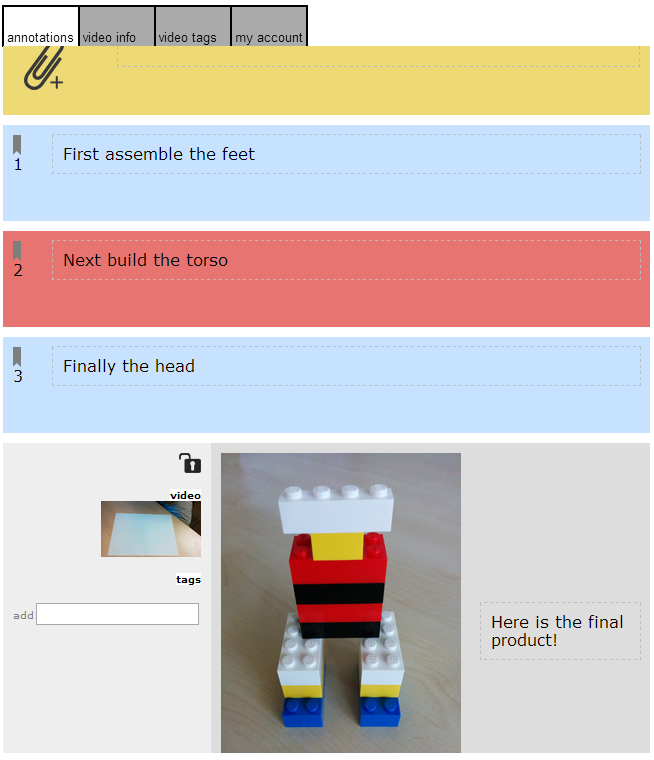} 
\end{tabular}
\caption{Editing bookmarks and annotations in a markup view. The user can add titles to bookmarks and create annotations. She can also drag media into any annotation (left). Media are re-sized and synced automatically (right).}
\label{fig:annotationsEdit}
\end{figure}

To create annotations users modify edit regions (yellow in Figure \ref{fig:annotationsEdit}) and click a commit button. 
After commit,  annotations are shown as grey rectangles in the markup panel. When needed, users can add media, including images, PDFs, video, links, audio, to an annotation via drag-and-drop (Figure \ref{fig:annotationsEdit}). Crucially, the authoring tool automatically lays out each inserted element -- no coding is required from the user (Figure \ref{fig:annotationsEdit}, bottom). When the user adds a PDF, the tool automatically generates a thumbnail and links it to the original; for audio and local video the tool generates the appropriate HTML5 control; for links to external video (e.g., YouTube) the tool automatically inserts the appropriate third-party iFrame controller.

Additionally, bookmarks and their associated video clip can be added as an annotation to any other video. To do this, the user grabs a bookmark from the timeline and drags it to a target annotation. The tool automatically detects this action and inserts the bookmark into an annotation in the markup panel. When dropped, the annotation lays out the bookmark using the HTML5 video element to play just the segment corresponding to the dragged bookmark. Figure \ref{fig:annotationsEdit} bottom right shows an annotation with a video bookmark from another tutorial and a photo.

\begin{figure}[tb]
\centering
\begin{tabular}{c} 
 \includegraphics[width=4.5in]{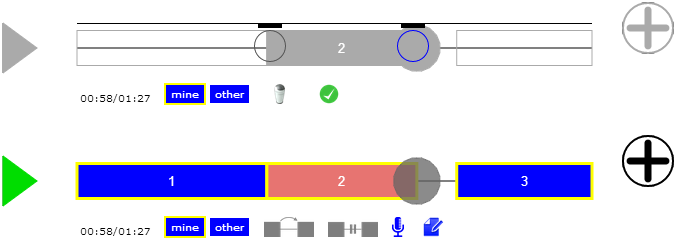} 
\end{tabular}
\caption{Editing bookmark segments on the timeline (top). The user can adjust the start and end times of any bookmark. Buttons below the timeline control how the player behaves when it reaches the end of a bookmark segment (skip or pause). Another button allows the author to re-record the audio track.}
\label{fig:timelineEdit}
\end{figure}

As mentioned above, ShowHow automatically creates one bookmark for each video clip in the tutorial. However, the user can also add a bookmark by clicking a ``+'' button next to the timeline which creates a bookmark at the current time in the video. 
Bookmark timings can be edited (using a modified version of the noUiSlider widget\footnote{http://refreshless.com/nouislider/}) or removed from the timeline (Figure \ref{fig:timelineEdit}). Users can only edit bookmarks that they create. 

In the timeline, the video's behavior at the end of a bookmark can be controlled. Users can click one  of the gray buttons at button of Figure \ref{fig:timelineEdit} to cause the system to pause at the end of each bookmark segment. This allows users following along with a how-to to take more time to complete a step. Another button causes the player to skip content between bookmark segments. This acts like a ``clipping'' tool, but it can be much more flexibly tailored since each user can decide which content is relevant. Video authors can configure the default setting for each of these options, but users can change them while viewing. 

Finally, our needfinding studies showed that how-to creators often felt their initial narration was distracted and confusing.  Thus, the interface allows authors to re-record the audio track of a video using an HTML5 media recorder. The system again uses a combination of {\tt ffmpeg} and {\tt MP4Box} to splice the audio and generate a new video.

\subsubsection{Making a document-based how-to with ShowHow}

\begin{figure}[htb]
\centering
\begin{tabular}{c} 
 \includegraphics[width=2.5in]{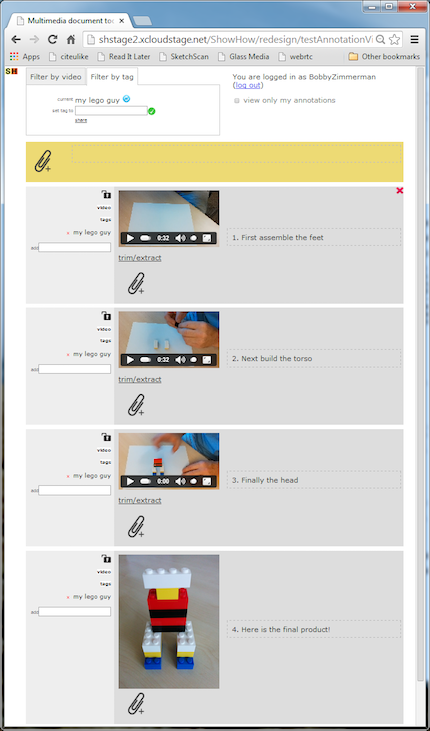} 
\end{tabular}
\caption{Representing the same content as in Figure \ref{fig:videoEdit} with a document-based approach. Instead of combining the three original clips into a video, the author created a new ``document'', created three separate annotations, one for each step, and dropped corresponding clip into its annotation. The author then added a final annotation containing a photo. The layout was arranged automatically. }
\label{fig:mmDoc}
\end{figure}

\begin{figure}[tbh]
\centering
\begin{tabular}{c} 
 \includegraphics[width=4.5in]{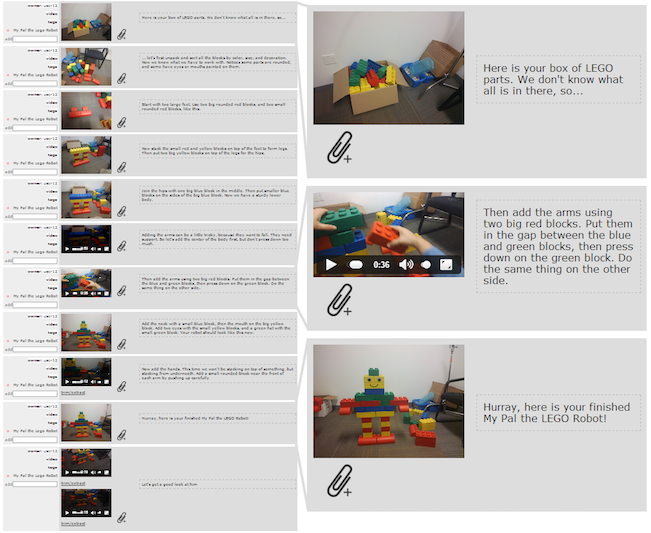} 
\end{tabular}
\caption{A document-based tutorial created by a participant in our authoring study. This 11-step document uses a combination of photos, videos, and rich text to illustrate the construction of a robot using large blocks. The first, seventh, and tenth steps are shown in detail (right). Note that in the seventh step the video starts at 36 seconds, indicating that the user trimmed the original video clip.}
\label{fig:mmDocExample}
\end{figure}

While the video-based approach relies on a composite video clip, the document-based approach lays out each step in the process separately using dedicated text and media.
This spatial approach has some advantages over the temporal approach offered by video: steps can be skimmed with a glance, searched more directly, and never need to be ``paused'' at step boundaries. Of course the approach has disadvantages as well -- since steps are not ``played,'' they need to be manually advanced and they typically do not include animations. We addressed this later issue by extending our drag-and-drop markup view.

Figures \ref{fig:mmDoc} and \ref{fig:mmDocExample} show a document-based tutorial created with ShowHow. A user can specify a tag and then begin creating associated annotations. Users can add any number of tags to each annotation. However, since annotations are ordered, a tag in the document-based tutorial view is functionally equivalent to a document name.  As in the video-based case, annotations include text and an arbitrary number of media elements. Each annotation is automatically assigned the current tag as it is created. Any number of tags can be applied to each annotation, promoting reuse. Furthermore, users can extract segments from uploaded videos to effectively clip longer uploaded videos. Figure  \ref{fig:mmDoc} shows how the same content from the video-based example can be compiled into a document-based tutorial.  The temporal sequencing of the video clips that comprise the video-based tutorial corresponds to the vertical arrangement of the multimedia annotations in the document-based tutorial. 

\subsection{Tutorial Access with Head-mounted Displays}
\label{sec:tutaccessdevice}

\begin{figure}[tbh]
\centering
\begin{tabular}{cc} 
\includegraphics[width=2in]{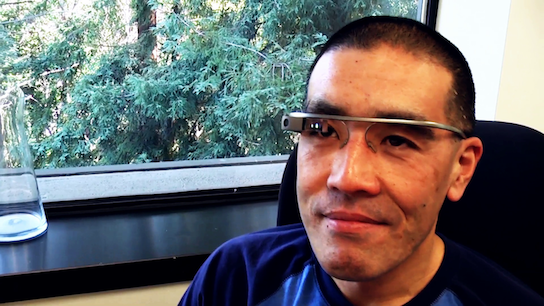} &  \includegraphics[width=2in]{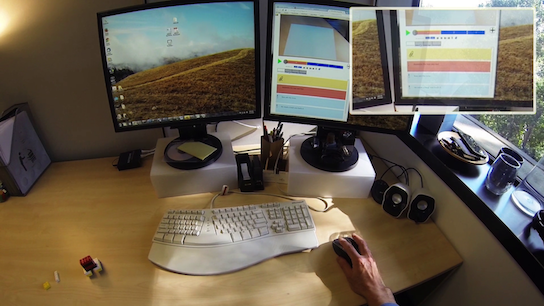} \\  
a & b \\
\includegraphics[width=2in]{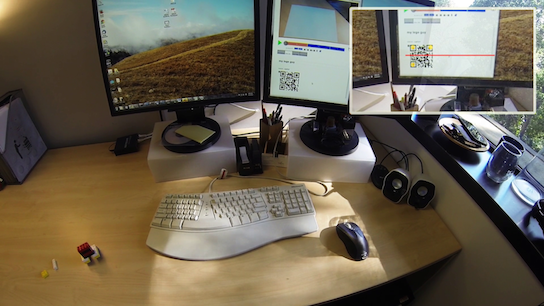} & \includegraphics[width=2in]{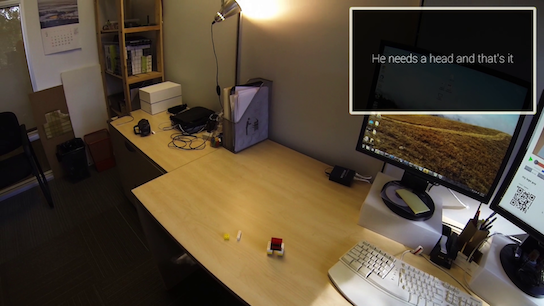} \\
c & d \\
\includegraphics[width=2in]{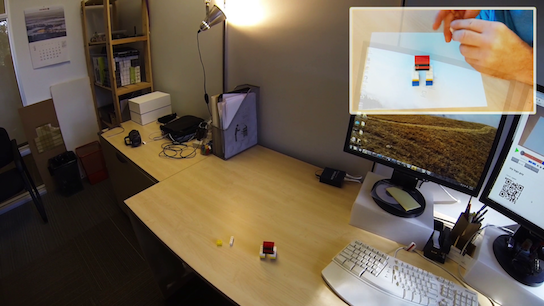} &  \includegraphics[width=2in]{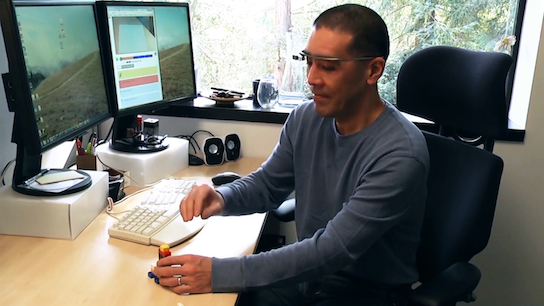} \\
 e & f \\
\end{tabular}
\caption{(a) A ShowHow Glass access user. (b) The user is viewing a ShowHow tutorial on the web. (c) With the Glass app (upper right of thumbnail) the user can scan the barcode associated with that tutorial. (d) The ShowHow app retrieves the document and loads all of the titles and media. The user can swipe Glass to move between steps (e), and interact with videos by tapping to pause and play. Also, swiping two fingers fast forwards or rewinds the video by 10 seconds. (f) Once complete, the user can replay the tutorial or scan a new one from the web. }
\label{fig:showhow_glass_access}
\end{figure}

We developed a Glassware app for accessing ShowHow tutorials. With this app users can scan a barcode associated with each tutorial to download and view that tutorial's media and meta-data (see Figure \ref{fig:showhow_glass_access}).
The access application exploits the tutorial's structure created in the process described previously. Each annotation or step is converted to a multimedia card stack. The text of the annotation comprises the top card in the stack. If the annotation includes only one media element (video or image), then the annotation will have only a single card, and the media element is shown behind the annotation text (for video-based tutorials the first media element is always the video segment linked to the annotation). In this case, tapping the trackpad while viewing the top-level text card toggles the visibility of the text. If the annotation includes more than one media element, tapping on the text card loads a new set of multimedia cards, one for each media element. From there, the user can swipe side-to-side to move between media cards or swipe down (go ``back'') to return to the top-level text card. Swiping side-to-side on  top-level text cards transitions between steps.

Video cards have more controls. Tapping on a video card alternately plays and pauses the video (when a video card is also a top-level card, the initial tap dismisses the annotation text and starts the video; the text stays hidden until the end of the video is reached or the step is reloaded). Also, swiping two fingers side-to-side on a video card forwards or rewinds the video by 10 seconds. Finally, in some cases an annotation or step uses only a subset of a video. In this case, the interface automatically seeks forward to the specified start time and stops the video at the specified end time.


\begin{figure}[tbh]
\centering
\begin{tabular}{c} 
 \includegraphics[width=3in]{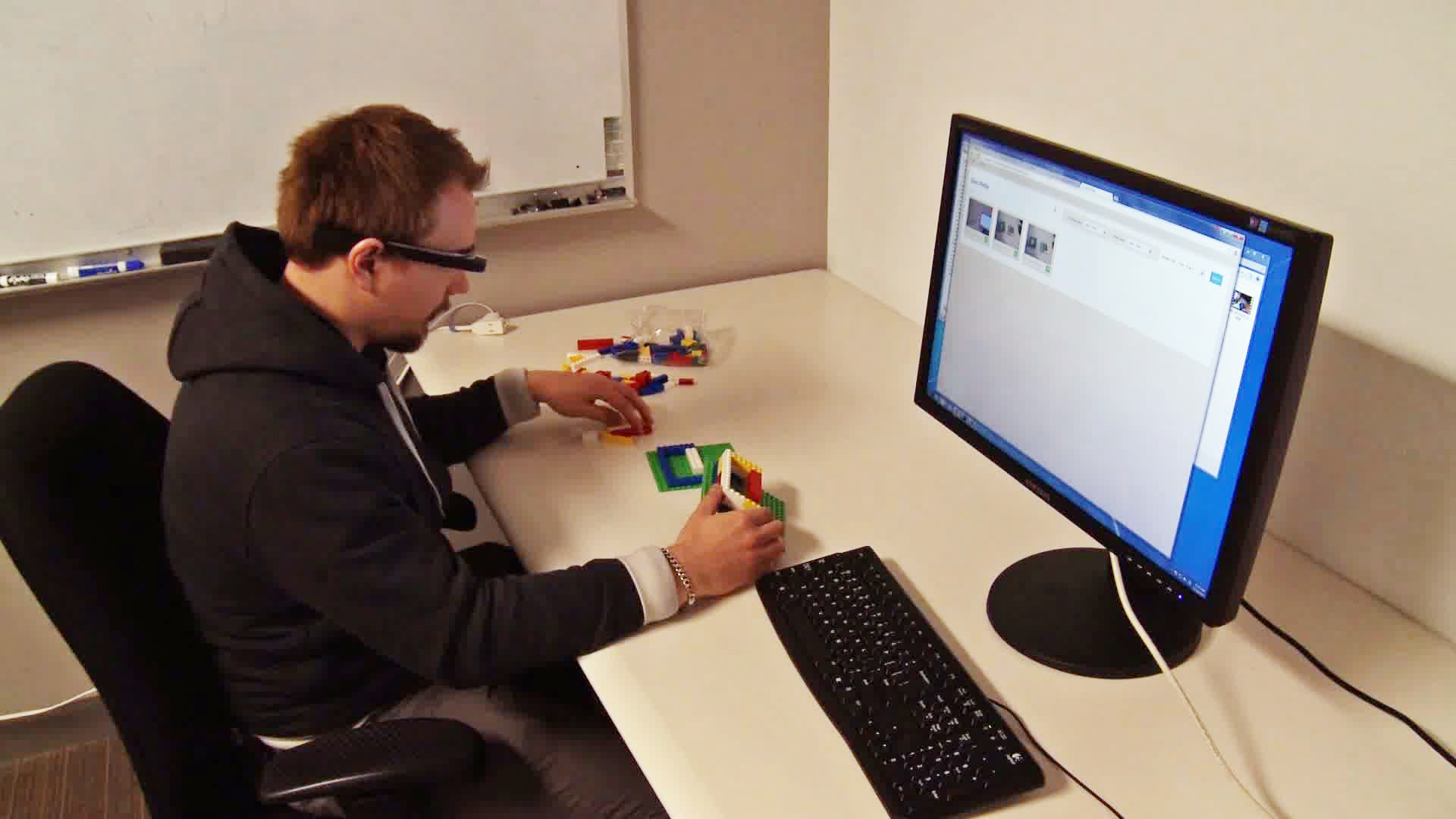} \\
 \includegraphics[width=3in]{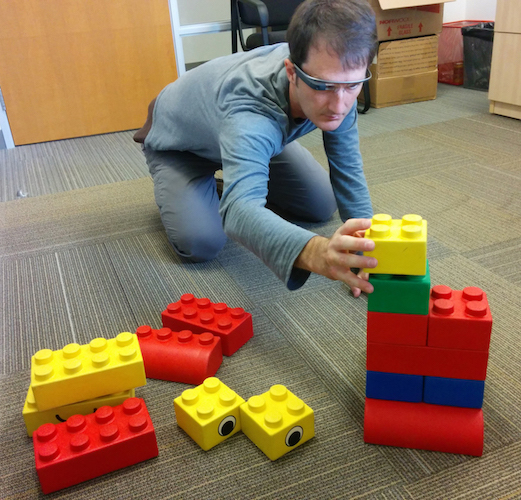} \\
\end{tabular}
\caption{A user completing the table-top task while recording with Glass (top). Media is synchronized automatically to the web page shown on the monitor. For the floor task participants used foam blocks much larger than their standard Lego counterparts (bottom).}
\label{fig:legoCompare}
\end{figure}


\section{Capture and authoring user Study}

To evaluate different capture methods, we compared portable cameras on tripod mounts with a head-mounted capture system (Google Glass with our capture application). Furthermore, we varied the capture location, table-top versus floor, to gain an understanding of how differing spatial requirements influence content creation. 
To assess the effects of the tutorial representation in the authoring tool, we had each participant author both a document-based tutorial and a video-based tutorial.

In this paper the focus of the user study is on how the authoring method influences tutorial creation strategies that in turn shape the media captured. Here, we perform a complete analysis of the users' strategies and the resulting tutorials' composition with regards to capture device, recording location and authoring tool. In \cite{carter2015creating}, we discussed the capture devices' impact on media included in the final tutorials.  For completeness, we include a brief review of selected findings from that work below. Although some results are by necessity overlapping, the main findings of the present study was not reported in \cite{carter2015creating}.


\subsection{Participants}

In total, 12 people (8 men and 4 women) 30-56 years old (M=43.5, SD=8.49) participated. All use smart phones and have recorded video with them, on average 6-12 times per year. All had experience with video editing, and half edited video more frequently than once a year. Ten participants were infrequent producers of tutorials, making tutorials less than once a year (the least frequent alternative). The remaining two made between one and six tutorials per year. Two participants had not used Google Glass. Of the remaining, five had used Glass once, and five a few to several times. 


\subsection{Task and procedure}

The task was to create tutorials that demonstrate how to build a robot with Legos. Each participant completed two tutorials using different sized blocks in two different locations: regular sized pieces sitting at a table, and larger foam bricks on the floor (Figure \ref{fig:legoCompare}, right). 
A 2x4 foam brick measures approximately 19x9 cm compared to a 3.1x1.5 cm regular Lego brick. The order of location was counter-balanced over all participants.

The study spanned two sessions, one for each task. In each session, the participant was assigned one of two capture methods (camera on tripod or Glass) as well as one of two authoring methods (document- or video-based). The assigned capture device and authoring tools were counter-balanced over all participants. The participants were encouraged to use a process that felt natural to them, meaning they were
not required to record all media before starting the authoring process. Task order and conditions were counter-balanced. 
Participants completed the study with a maximum of two days between the sessions. The first session started with participants answering a background questionnaire.

Each session was divided into two phases. In the first phase, participants watched two short video tutorials describing the capture and authoring tools.
Next, they designed and built a Lego robot model while practicing taking photos and recording video. The participants were free to design the robot, with the only restriction that it should be at least eight layers tall. During this phase, participants were encouraged to try combining video clips and pictures into a tutorial. All of the capture and recording data was discarded before the second phase. 

In the second phase, participants were instructed to build the exact same robot as in the first phase using an identical set of Lego pieces, to record the building process, and to create the final tutorial with the authoring tool. 
While capturing media for the tutorial, participants were free to choose the style and focus of their tutorial. Narration was encouraged but not mandatory.
Participants were asked to frame the video and photos to best suit their needs by moving the tripod or changing recording position while wearing Google Glass. The participants were told that their final tutorial should be sufficient to enable someone to replicate their exact robot.


While recording in the floor condition, participants were asked not to move the robot, but instead to move around it with the camera as needed. We took this approach because in most cases the robot was not stable enough to easily move, and also because this better simulates how users would record larger objects (e.g., cars, home repair, etc.).
During the table-top condition, users were free to move the robot.
After each phase, the participants completed a questionnaire 
including both Likert-scaled as well as free-form questions about their experience with the capture and authoring tools used in that particular session. 
After both sessions were completed, participants filled out one additional questionnaire comparing their experience with the capture and authoring processes.
The length of the sessions varied among the participants, the shortest one being approximately 30 minutes and the longest ones around 140 minutes.

\subsection{Equipment}

The capture device was either the Google Glass running the ShowHow capture application or a smart phone, Nexus 4, on a tripod. Both devices allowed the user to take pictures and record video, and uploaded media automatically. The Nexus 4 synced media to a Dropbox folder. The media could be dragged from the Glass ShowHow web page or Dropbox folder into the authoring tool for further editing.

The authoring tool was either the ShowHow document-based or video-based tool. With the document-based tool, steps could contain any combination of pictures, videos, or text. In the video-based tool, the focus was on creating a composite video from several clips, however participants were free to add bookmarks and multimedia annotations to the video. 
A suggested length of the final video-based tutorial was 2-3 minutes, however no real restrictions were applied. 
Annotations and detailed bookmarks were encouraged but not required.
We ran the desktop applications on a Windows 7 PC with Google Chrome and Dropbox installed. We also recorded video of each participant completing each task with ceiling- and boom-mounted cameras.

\subsection{Data Analysis}
A complete within-subjects design would require each participants to make 12 tutorials. However, this was deemed as impractical since each each session required a significant amount of work and time for each participants. Yet, we were interested in the participants' comparison between the different capture and authoring methods. Hence we limited the number of tutorials each participants needed to make in the study to two, which allowed each participant to try out the two capture, authoring methods and locations to be able to compare the conditions. As a result, only main effects could be calculated as within subject repeated measurements, while interaction effect would by necessity be between-subject measurements. To simplify the analysis, we choose to regard the design as between-subject design when a full-factorial analysis was performed. In these cases, we use an ANOVA with the independent variables capture device (Glass vs. Camera+Tripod), authoring tool (Document-based vs. Video-based) and location (Table-top vs. Floor).

We performed a separate paired t-test to check for order effects. No order effect were found for session duration, average video duration, number of video clips, photos and annotation captured. The media required for making a comprehensive tutorial is not likely to vary depending on previous experience, but on the content and structure of the tutorial. 

\subsection{Tutorial quality ratings}

We asked three judges, who were professional and amateur photographers and videographers, to evaluate each tutorial on a variety of quality metrics, describing the quality of the media and the content.
All metrics were rated on a seven point Likert scale. Media quality was rated based on sharpness, focus and framing. The video and photo quality metrics were combined into a media quality metric since not all tutorials contained audio, video, or photos. It was calculated as the average of the scored metrics. Similarly, the various tutorial content metrics were combined into one tutorial quality measurement to simplify 
the analysis.

Judges collaboratively rated two tutorials to establish common ground for interpretation of the metrics. After rating all tutorials, the judges met to discuss and reach consensus on each metric. Consensus was defined as two of three judges reporting the same score with the third being at most one point below or above the rating of the other two judges. Finally, the median of the judges' ratings was used as the final score. 

 \begin{figure}[thb]
 \centering
 \begin{tabular}{c} 
  \includegraphics[width=3in]{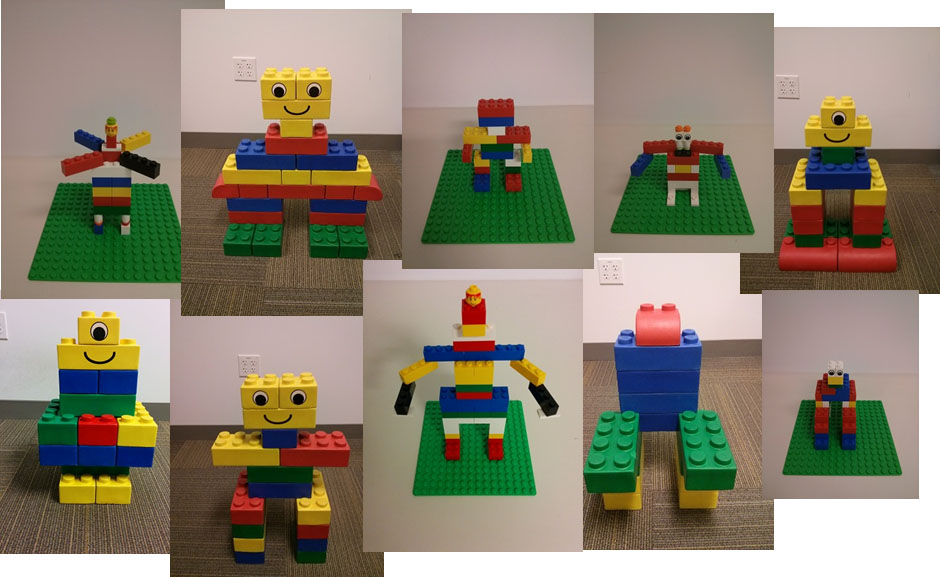} \\
 \end{tabular}
 \caption{Participants created a wide variety of robots.}
 \label{fig:bots}
 \end{figure}

\subsection{Results and discussion}


To explore how participants used the tools available to them, we examined the completed tutorials in terms of the media used and annotations created. Table \ref{tab:sumtable} summarizes the data we explore in this section. 
\subsubsection{Tutorial characteristics}

When working with the video-based tutorial tool, the participants were restricted to combine videos into a tutorial with photos as annotations. The resulting videos often depicted the entire building process (11/12 tutorials). Figure \ref{fig:videoExample} shows a common structure for a video-based tutorial: a video with bookmarked steps and photos showing the final robot. Photos added to the video-based tutorials showed the completed robot (6/12), and occasionally showed a partially built robot (2/12), or materials used (2/12).  In contrast, when participants authored a document-based tutorial,  the video clips were more focused: illustrating difficult steps (5/12), specific pieces used (3/12), or showing the robot from multiple angles (4/12). Only three of twelve document-based tutorials included the whole building process in video clips. These tutorials used photos to describe the various steps.  Figure \ref{fig:mmDocExample} illustrates a common structure for a document-based tutorial in which a combination of photos and short video clips outlines the building process in discrete steps. The video in the seventh step shows a difficult step in the building process. These observed differences were confirmed as we more closely examined the different variables summarized in Table \ref{tab:sumtable}. 

\subsubsection{Tutorial Creation Duration}
The amount of time participants spent preparing, recording and authoring the tutorials (phase one and phase two) varied greatly. On average, participants spent 23 minutes (SD=15.0), but some 
used as little as 11 minutes while others spent up to one hour preparing, recording and 
authoring the tutorial. We did not find any main effects on capture device, location, or authoring tool, however we found a 
significant interaction between capture device and authoring tool ($F(1, 16)=7.408, p < 0.05$). When participants used the tripod set up and the video-based authoring tool, the session duration was significantly longer than when they used the document-based tool (Tripod + Document: M=19 min , SD=4.9; Tripod + Video: M=54 min, SD=31.4; Tukey HSD, $p < 0.05$). No other differences were significant. One possible explanation is that in tripod+video condition more short video clips were recorded so longer time was needed for recording and set up of the recording environment compared to the Glass+document condition. 

\begin{figure}[thb]
 \centering
 \begin{tabular}{c} 
  \includegraphics[width=4.5 in]{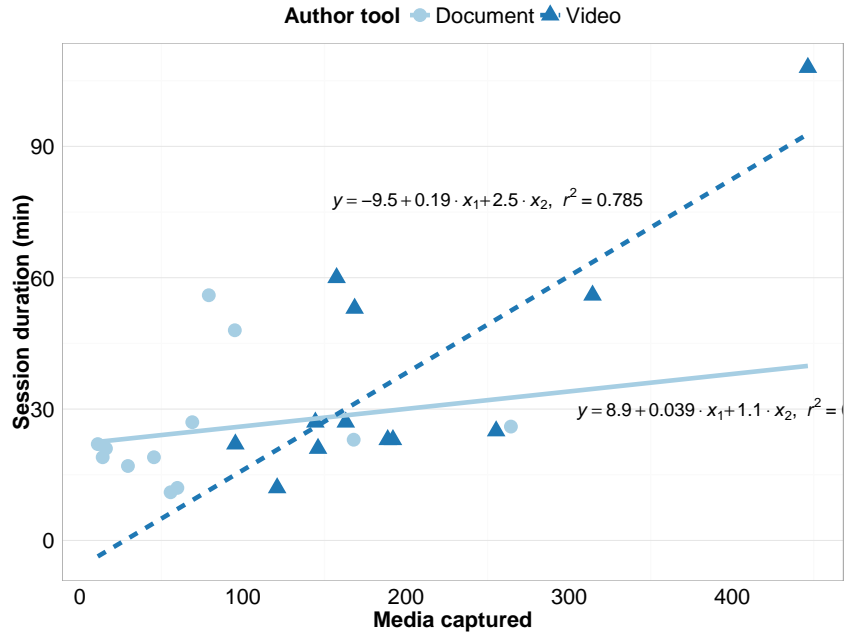} \\
 \end{tabular}
 \caption{Average number of video clips per tutorial for each condition. Error-bars represent $\pm 1$ standard error. }
 \label{fig:sessiondur}
 \end{figure}

The session durations were highly influenced by how much media the participants captured (Figure ~\ref{fig:sessiondur}). Interestingly, we found that when participants used the video-based authoring tool the tutorial session duration could be predicted by the total video duration recorded ($\beta=0.19, p<0.001$) and the number photos collected ($\beta=2.6, p<0.05; F(2,9)=16.45, p<0.001, R^2=0.785$). When participants used the document-based authoring tool, the number of photos was a significant predictor ($\beta=0.11, p<0.001$), while video duration was approaching significance ($\beta=0.039, p=0.0756; F(2, 9)=24.53, p<0.001, R^2=0.845$). The two linear regressions for the video- and document-based authoring tools were significantly different ($F(9, 9)=5.486, p<0.05$). As the slope in Figure ~\ref{fig:sessiondur} suggests, the cost of capturing more media when using the video-based authoring tool is higher than when using a document-based authoring tool.

\begin{table}[t]
\tbl{Summary of media used in tutorials for each of the conditions: average number video clips, photos and annotations, as well as the average of video clip duration (s). Standard deviation within parenthesis, $^*$ significant at 0.05-level,  $^{**}$ significant at 0.01-level.}{%
\begin{tabular}{c|c||c|c|c|c|c}
Condition &	Levels 
&	Session duration (min)	
& \# clips &	Dur./clip (s) &	\# photos & \# annotations \\ \hline \hline
All	&               &	
23 (15.0) &	
3.4 (2.68) & 59 (54.7)	& 5.0 (5.66) &	7.5  (3.7) \\ \hline
\multirow{2}{*}{Capture}	& Glass     &	
27 (12.9) &
2.8 (2.69) & 81 (67.5)$^{**}$ &	5.5 (7.03) &    7.5 (4.05) \\ 
	& Tripod        &	
	30 (17.6) &
	4.0 (2.66) & 35 (20.2)$^{**}$ &	4.6 (4.14) &	7.5 (3.53) \\ \hline
\multirow{2}{*}{Location} &	Floor   &	
29 (17.0) &
2.7 (2.14) & 65 (65.6)        &	6.5 (7.14) &	8.1 (3.87) \\ 
 &	Tabletop        &	
 27 (13.2) & 
 4.2 (3.04) & 52 (42.2)        &	3.8 (3.59) &	6.9 (3.63) \\ \hline
\multirow{2}{*}{Authoring} &	Document &	
25 (13.6) &
2.8 (2.52) & 27 (17.6)$^{***}$ &	8.1 (6.50)$^*$ &	8.1 (3.96) \\ 
 &	Video           &	
 32 (16.4) &
 4.0 (2.83) & 83 (61.1)$^{***}$ &	2.0 (2.13)$^*$ & 6.9 (3.53) 
\end{tabular}}
\label{tab:sumtable}
\end{table}

\subsubsection{Video usage}
Carter {\em et al.}~\cite{carter2015creating} reported how participants captured video and photos with the two different capture devices at the two different locations and found no difference depending on capture method. Here, our focus is on how the author tool affects the choices participants make when selecting to capture photo and videos. 

\begin{figure}[thb]
 \centering
 \begin{tabular}{cc} 
  \includegraphics[width=2.52 in]{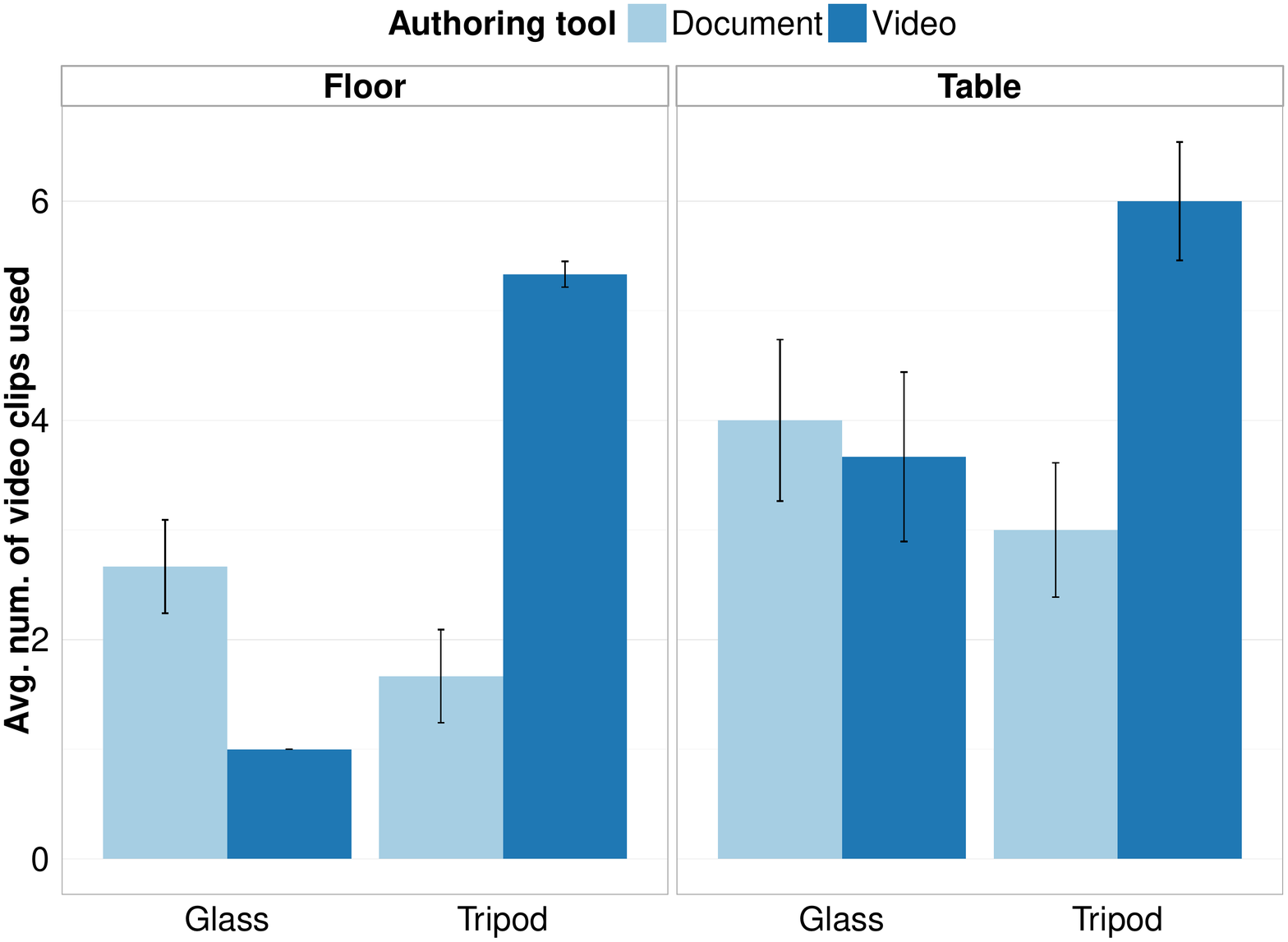} &
    \includegraphics[width=2.52 in]{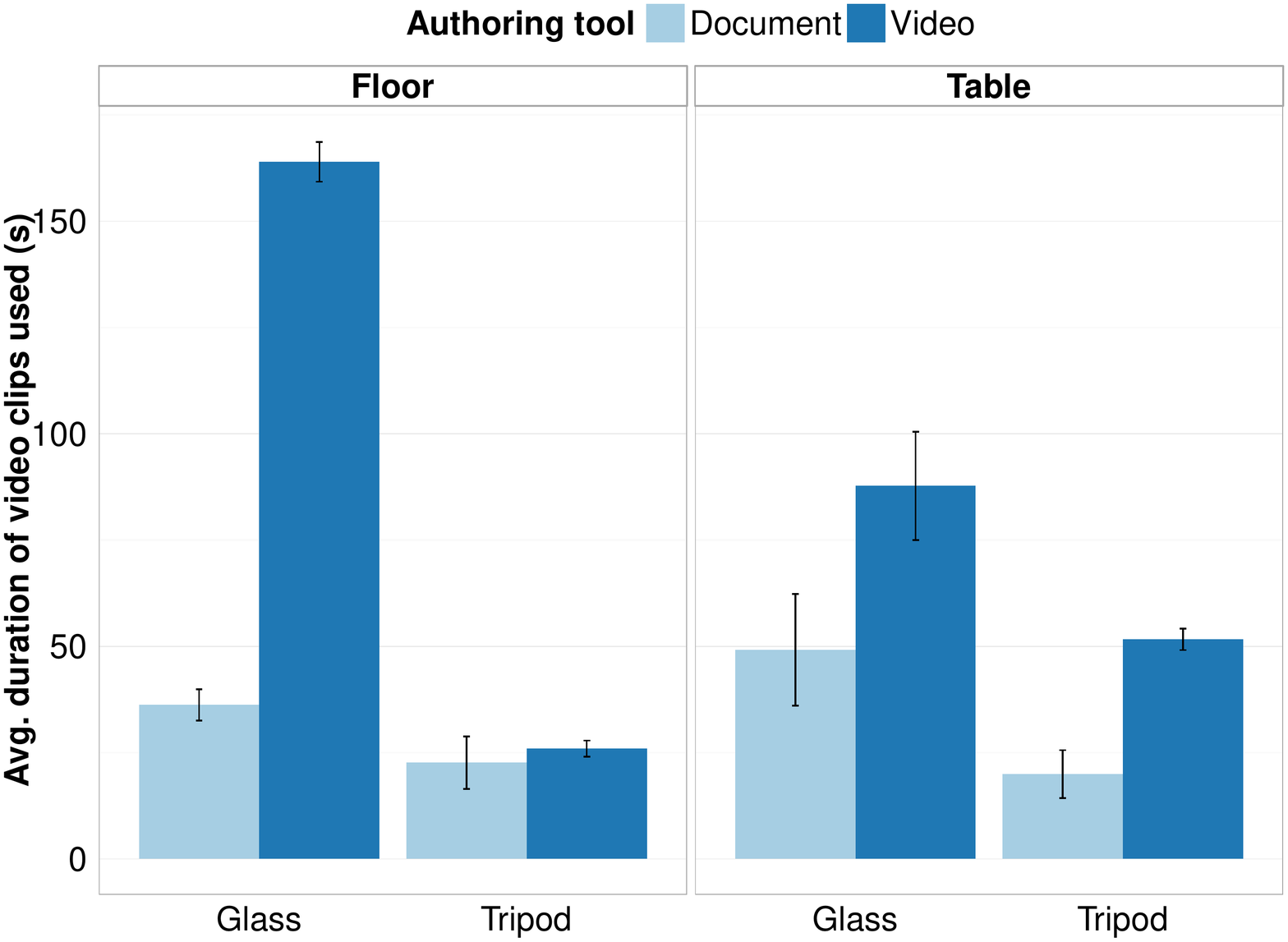} \\
  (a)  & (b)
 \end{tabular}
 \caption{Left: Average number of video clips per tutorial for each condition. Right Average duration (s) of video clips per tutorial for each condition. Error-bars represent $\pm 1$ standard error.  }
 \label{fig:num_clips}
 \label{fig:session_duration}
 \end{figure}

Participants had to include at least one video in their video-based tutorials, but none were required in their document-based tutorials. 25\% of the document-based tutorials (3 of 12) did not contain any video clips. Overall, the average number of video clips in the video-based tutorials was higher (4.0 vs. 2.8 video clips), but this difference was not significant ($F(1, 16)=1.248$). 
Figure \ref{fig:num_clips}(a) shows the number of video clips used in the different conditions. 
We found that the capture method as well as the authoring method affected the number of video clips captured and used. All participants using Glass with the video-based authoring tool only recorded and used \emph{one} video in their tutorial. Also, when participants used a camera on a tripod in combination with video-based authoring, they included more video clips, on average 5.7 (SD=1.75), in their final tutorials than in all other conditions (see Figure \ref{fig:num_clips}). Overall, we found a near significant interaction between authoring tool and capture device regarding number of video clips used ($F(1, 16)=4.306, p=0.0545$), which supports our observations above. However, we did not find any main effects on the conditions capture device, capture location and authoring tool.


We also found that the average video clip duration was highly effected by author tool in combination to capture device. 
Figure \ref{fig:session_duration}(b) illustrates how the average duration varied with the different conditions. When participants used the video-based authoring tool, the video clips were on average 56 sec longer than when using the document-based tool. This difference was significant, $F(1, 13)=20.597, p < 0.001$.  In addition, when participants used Glass, they recorded on average 45 sec longer video clips than when using camera+tripod ($F(1, 13)=12.031, p < 0.01$). Our analysis also revealed a significant interaction between capture device and location ($F(1, 13)=4.693, p < 0.05$) and between capture device and authoring tool ($F(1, 13)=12.150, p < 0.01$). 


When using Glass and the video-based authoring tool, participants' video clips were on average significantly longer compared to all other combinations of capture device and authoring tool, mainly because they recorded their whole building process in one shot as previously discussed (Tukey HSD, video-based vs.: video-based + tripod $p<0.01$, document-based + Glass $p<0.001$, document-based + tripod $p< 0.01$).  When participants created a robot on the floor with Glass as the capturing device, they recorded significantly longer video clips, on average 71 sec longer, than when they used a camera on a tripod (Tukey HSD, $p<0.001$). 
Our findings support the observation that all three conditions, authoring tool, capture device and location, influence the choices the participants made during tutorial creation with regards to video recording. 


These results indicate that with Glass as the recording device during tutorial creation, the authors felt confident enough that the recording would capture the building process that they let the camera roll and focused on showing how to create their robot. Based on participants’ behavior, this kind of recording method is well supported by the video-based authoring tool. With a tripod on the other hand, our data suggests that the authors created shorter snippets, checking the recording of a step and adjusting framing before moving on to record the next step. 

 
 
\subsubsection{Photo \& annotation usage}
Participants could add photos to annotations in both their document- and video-based tutorials. As noted, some participants elected to use only photos when working with the document-based authoring tool. We found that when creating a document-based tutorial, participants included on average four times as many photos compared to their video-based tutorials ($F(1, 16)=8.161, p<0.05$). The capture device and location had no measurable effect on the number of photos included.

Annotating and bookmarking video content are integral capabilities of the ShowHow toolset. Both features allow the user to create steps in the tutorial. Since these features are functionally similar in the final tutorials, we added the number of bookmarks to the number of annotations in the video-based authoring condition and compared the sum with the number of annotations in the document-based authoring tool. On average, the participants created 7.5 (SD=3.7) annotations in all conditions. We found no difference depending on capture device, authoring tool or location. 

The characteristics of the authoring tool appear to change how the user approaches the capture process. 
The document-based tutorial creation tool emphasizes video capture of more focused steps in the process, and users take more photos to support step-by-step instructions.
Next, we look closer at participants' experience using the capture devices and authoring tools.


\subsubsection{Head-mounted vs. standard capture}

\cite{carter2015creating} 
reported that participants valued the convenience and confidence that head-mounted capture with Glass gave them despite potential degradations in content quality. This trade-off was particularly evident when working with the larger model that required movements over a larger area. Setup with Glass is minimal: the participants can simply move to a new location. Feedback on how well the content is framed is accessible with a glance. With camera on tripod, the participants needed to check the framing and focus for every new location and it was not always obvious when the participants move out of the camera frame while working. Checking the framing while building the Lego model can be disruptive to the main objective of show-and-tell. The number of framing shifts performed were higher with Glass than for camera on tripod, simply because framing shifts can be achieved with minimal efforts with Glass. The participants simply needed to move their head to better convey important aspects of the building process. With Glass, the video can become more unstable than with a camera on tripod, but the value of unrestricted movement and instant feedback was clearly higher for our participants. When working at a tabletop, using camera on a tripod was considered a good alternative since the camera could provide easy zoom to show small details and a stable recording. Small models can easily be moved around for showing important aspects of the building process.

\subsubsection{Authoring tools}

Our participants preferred the document-based authoring tool to the video-based authoring tool in multiple situations. 
Eight of twelve participants describe this tool as easy, natural, or intuitive to use. In comparison, the video-based tool was only described as easy by two participants. A common theme 
was that the structure of the document-based authoring tool fit well with tutorials:

\begin{quote}
Document based tool gave me more freedom to record steps (structurally) using photos + videos. Basically I could shoot surplus photos and videos and then choose from among them to describe steps. It was slightly more intuitive than creating bookmarks and annotating them. (Participant 6)
\end{quote}

\begin{quote}
It was surprisingly easy and intuitive to use. The document structure seems to work well for how-to videos, as we basically want to show a sequence of build steps. (Participant 8)
\end{quote}

Another common view was that the document-based authoring tool was more flexible and potentially more useful than the video-based authoring tool. One participant (Participant 7) said: ``I liked being able to include images and text as an integral part of the final product. I felt like this final product might actually be usable for someone else trying to learn from it.''

In contrast, participants reported that the video-based authoring tool had a longer learning curve (4 of 12) and that it was somewhat frustrating to use (3 of 12). Also, some of the participants (3/12) found the authoring tool inflexible because it forced them to record more video clips than required for making a comprehensible tutorial. One of them, Participant 9, said: ``I felt a bit constrained to take video for things that might not need video.  End result is a video with a bit of annotation - no surprise.'' However, participants pointed out one of the strengths of the video-based authoring tool is that it uses a familiar structure: ``Video consumption is more familiar to the masses and works well on many devices with different form factors.'' (Participant 2)

As also shown above, the structure of the authoring tool prompted participants to capture more photos and to record and use video clips of only selected parts of the building process. This interaction between the authoring tool and media capture was evident in the participants' comments:

\begin{quote}
It prompted me to take more photos to add emphasis on certain stages of the build process (Participant 2, about document-based authoring tool)
\end{quote}


\begin{quote}
I did not shoot stills while I was building for my own convenience, but if I had, I think it might have been awkward to insert them anyway. (Participant 12, about video-based authoring tool)
\end{quote}

\subsubsection{Tutorial quality}
The tutorials were rated on media quality and content quality. 
Of the top ten rated tutorials on tutorial quality, seven used the document-based format and were recorded in the floor condition. The combination of video and photos that was frequently used in the document-based tutorials appears to improve the quality of these tutorials. These results also match the participants' feedback that the document-based authoring tool improved the quality of the tutorials.

Of the ten top rated tutorials on media quality, seven were recorded with Glass and three with camera on tripod. In comparison, of the ten tutorials with the lowest ratings, four were recorded with Glass and six were recorded in the tripod condition. This is notable since participants report that the recording quality was better when using camera on a tripod. A head-mounted camera is more sensitive to head movements and the camera angle on Google Glass is not quite designed for recording manipulations with the hands close to the body. For this kind of recording, participants needed to consciously frame the capture. However after authoring the tutorial, the participants managed to create tutorials with equal or better quality by using Glass than by using a camera on a tripod.

Tutorial and media quality was not effected by session duration; instead the tutorial quality was affected by the number of video included ($\beta=-0.044, p=0.013$), the ratio of video kept after editing ($\beta=0.341, p=0.063$) and the total number of annotation and bookmarks created ($\beta=0.823, p<0.001$; Multiple linear regression, $F(3, 30)=7.717, p<0.01, R^2=0.536$). Note that number of videos have a negative $\beta$, meaning that fewer video resulted in higher quality. These results indicate that the effort the participants put into creating the tutorial, adding bookmarks and annotation, and editing the video clips for the tutorial helped to improve the overall quality of them independent of capture device, location or authoring tool.



\subsection{Conclusions and Discussion}
Our participants clearly preferred using a head-mounted device for capturing building processes involving larger objects or spatial areas. On the other hand, while working on a desktop, the participants preferred the camera and tripod. Overall, head-mounted capture was considered at least as effective as a standalone camera on a tripod for tutorial content capture, both by our participants and by judges comparing media quality.

Our results point to an interaction between capture device, work location, and authoring tool. When using head-mounted capture and video-based authoring, the participants only recorded one video clip of the whole building process. When using the document-based authoring tool, participants captured more photos and shorter video clips. Our participants preferred the document-based authoring tool since they found it more naturally integrated different media and supported the character of tutorial with separate steps. For participants mixing photos and video clips in their tutorial, we found that the total time spent on creating tutorials was less for document-based tutorials.

The goal with our capture and authoring tools was to lower the barriers for tutorial creation. For this to be accomplished, the time and effort the authors need to put for creating a good quality tutorial needs to decrease. Our study shows that with higher ratio of mixed media capture, authors reduced the session duration. Yet, the session duration did not impact the tutorial quality. Instead, tutorials with fewer well edited videos and more annotations and bookmarks increased the tutorial quality. Document-based tutorials had more focused video, more mixed-multimedia capture, and more annotations. Our results indicate that authors not only preferred creating document-based multimedia tutorials, but that this method can streamline content creation while retaining good quality.


\section{Tutorial access user study}
In the capture and authoring study, we showed that participants preferred using a head-mounted device to capture media for their tutorials since the device seamlessly recorded their process. Intuitively, head-mounted displays have several useful attributes for accessing tutorials; they are handsfree and include a glanceable display always accessible to the user. Previous research using head-mounted display for video-mediated communication have shown conflicting results on its usefulness. \cite{Fussell:2003} used a head-mounted camera to record the actions of a worker while a remote helper viewed the video stream. They found that head-mounted capture produced poor quality content with quick camera shifts and shaky video, in turn impeding the helper in understanding the video. A recent study, \cite{Johnson:2015} used Google Glass as the capture device in a similar scenario to Fussell et al., and compared it with capture using a tablet. Johnson et al. found that in situations requiring the helper to move between different tables, the head-mounted capture allowed the helper to provide more directing commands and proactive assistance. Although these are interesting results, accessing a tutorial is fundamentally different to real-time computer-mediated communication since the communication is one-way. Tutorial viewers have to solve their problem using the composed tutorials. Video-recordings' poor quality cannot be compensated for by verbal utterances as during real-time communication. Tutorial access has received little attention from the HCI community in relation to real-time computer-mediated communication. In this study, we were in particular interested in users' experience with a head-mounted display for content access, in particular in light of the tutorial creators' preference for head-mounted capture.

One notable consideration is the impact of the recording perspective for understanding recorded content. A first-person perspective recorded from the eyes of an instructor can be beneficial for accessing tutorials since it would reduce the need to perform mental rotation, a mentally demanding task \cite{shepard1971mental}. When working with a large physical object, a user cannot simply turn around the object to achieve the same perspective as the recorded video instructions but needs to move around the object. If users feel a need to align to the recording perspective, would a head-mounted access device give users a greater freedom to choose a building position and does it impact their performance?

To assess how well head-mounted displays worked from the point of view of a tutorial consumer, we designed a study with within-subject design that compared tutorials captured in first person perspective using head-mounted capture with tutorials captured using a camera on tripod from a third person perspective. In addition, we were interested to see if the access device affected the utility and experience of the tutorial. Hence, the study had two independent variables (recording perspective and access device) with two levels each. All participants were exposed to all four conditions and the order of the conditions were counter-balanced over all participants.

\subsection{Participants}
We recruited 19 participants for the study, and we used data from 16 participants (4 women and 12 men). Ages ranged from 23 to 50 years, median age was 41 years. Three participants were excluded since ShowHow Glass crashed during the study session. The version of Glass used in the study tended to become unresponsive when it became hot, and running video over longer time periods consumes enough power to relatively quickly heat up Glass.

Most of the participants had used Google Glass before (Once: 6, A few times: 7, several times: 2), only one participant had not used Glass before the study. However, most participants were novices with Glass, only two participants had used Glass more than a few times. The participants were frequent users of tablets, only one participant had never used a tablet, while 12 used tablets more than once a month. The remaining four participants had used tablet a few times.
Lego building frequency ranged considerably within the participant group. 
But the participants had a fairly high confidence that they could follow a regular Lego instruction with a mean rating 5.4 (Median=6, SD=1.26), on a scale from one (not at all confident) to seven (very confident).

\subsection{Equipment and Tutorials}

To display the tutorials, we used Google Glass as the head-mounted display. It ran the tutorial access application described in \ref{sec:tutaccessdevice}.
We also made a version of ShowHow Glass for the Nexus 7 tablet. The user interface was the same as for ShowHow Glass, and the interactions were mirrored on the tablet. For instance, a single tap on the touch screen started or stopped a video. There were two exceptions. First, the tablet user could simply click the device's back button rather than swiping down as they would on Glass. Second, the direction of the side-to-side swipes (used for moving between steps and seeking videos) were reversed on the tablet, reflecting a strong user preference that arose in piloting.

We created four tutorials for this study. We considered using the tutorials created in the capture and authoring study, however, the quality and complexity of these tutorials varied greatly so it was not possible to find enough tutorials of similar quality and complexity. The media for the tutorials was captured by Glass or Nexus 5 on a tripod during the same recording session. This procedure ensured that both recordings contained the same verbal and non-verbal instructions. Care was taken so that only the arms of the instructor appeared in both recordings.

The camera on tripod was positioned to capture a third-person perspective of the building area. During tutorial creation, the direction of both recordings in relation to the model was noted. This was used as a reference point to  compare where participants positioned themselves in relation to the model and recording angle.

Each tutorial contained both video and photos of a Lego model using the large Lego block in the capture and authoring study. Three tutorials had five steps and one had four steps. All tutorials were tested to make sure they had the same level of complexity before the study.
The models created for the tutorials were a dragon, fort, robot and race track. Two of these models, fort and race track, had a rectangular shape where the two ends needed to match.
We compared the tutorials' quality based on participants' ratings. We could not find any significant difference between the four tutorials. Further, no significant differences were found on tutorial quality depending on recording device.

\subsection{Task and Procedure}
In this study, we compared two tutorial access devices, a Nexus 7 tablet and Google Glass, and two recording perspectives, first- and third-person recording. A first-person recording corresponds to recordings done with Glass, where the viewer has the same field of view as the tutor. A third-person perspective corresponds to the view of the recording environment that includes the activity and the tutorial creator. 
The participants were first exposed to one of the two access devices and completed two tutorials, one captured with Glass in first-person perspective and one with camera on tripod in third-person perspective, before moving on to the other access device. The order of the access device and recording-perspectives, as well as the four different tutorials were counter-balanced over all participants.

At the start of the study session, the participants were introduced to one of the access devices, either Google Glass or Nexus 7. They were given a short tutorial to try out and familiarize themselves with the ShowHow access application on the device. They could use as much time they needed to feel comfortable with the access application and device.

When the participants obtained enough experience with the access application, they were given one of two tutorials to view. The participants' task was to follow the instructions in the tutorial to build the model. The participants had access to a pile of large Lego blocks for construction. The pile included all Lego blocks needed with plenty of extra pieces. 

While the participants were building the model, the instructor took notes of the process. These notes included how the participant positioned themselves in relation to the model, the camera view and the tutorial instructor. 

When the model was completed, the participant answered a short questionnaire about the building process. Next, the participants were given the second tutorial. This tutorial had a different recording view compared to the first tutorial. After this tutorial was completed, the participants responded to the same short questionnaire and were introduced to the second access device. The procedure was now repeated with an introduction to the access device and completion of two tutorials.

After all four tutorials had been completed, the participants were asked to fill in a questionnaire asking them to compare their experience with the two access devices and the two viewpoints, first person and third person. The study session lasted about one hour and ten minutes.

\subsection{Data Analysis}
\begin{figure}[tbh]
 \begin{center}
 \begin{tabular}{cc} 
  \includegraphics[width=2.5in]{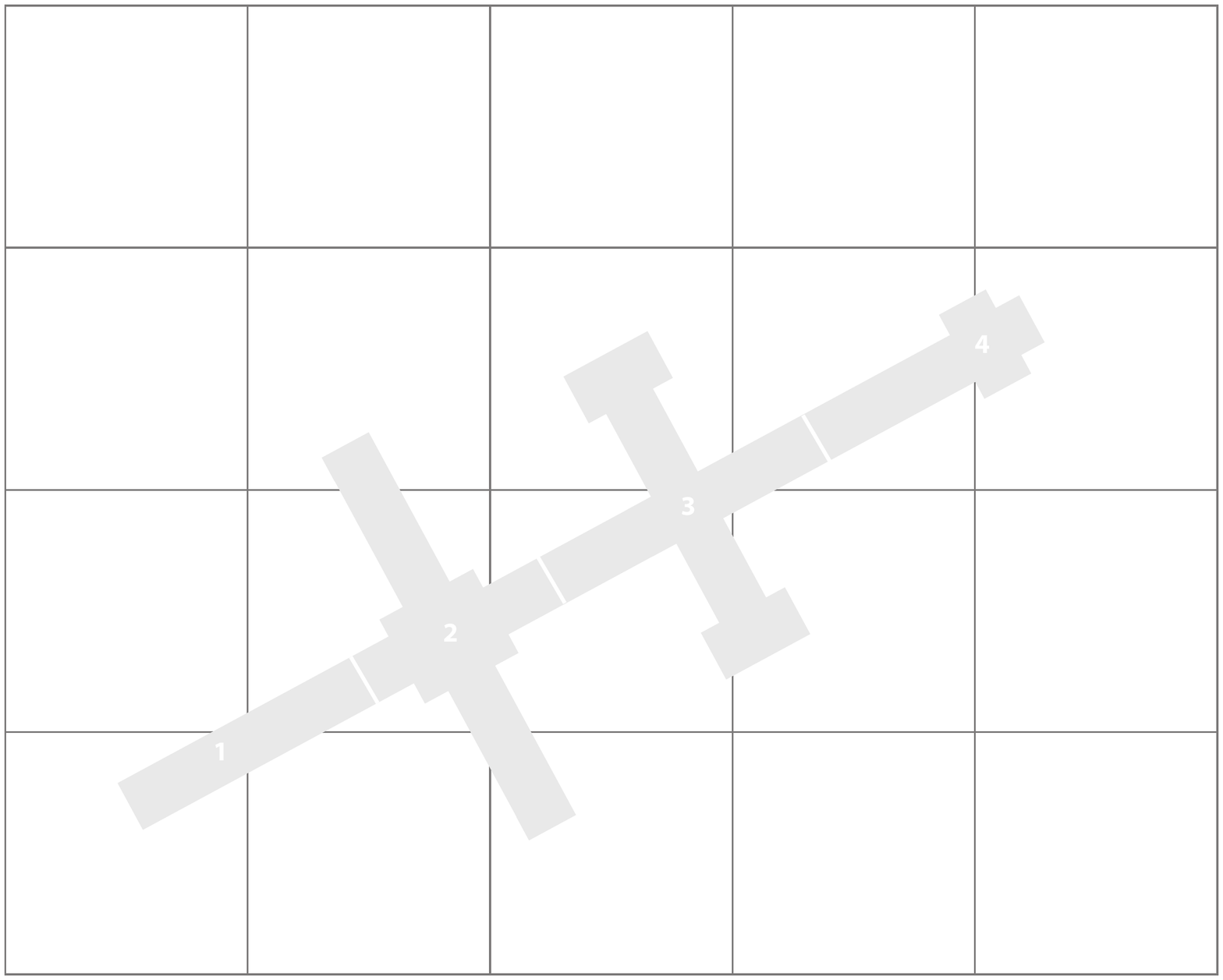} & 	\includegraphics[width=2.5in]{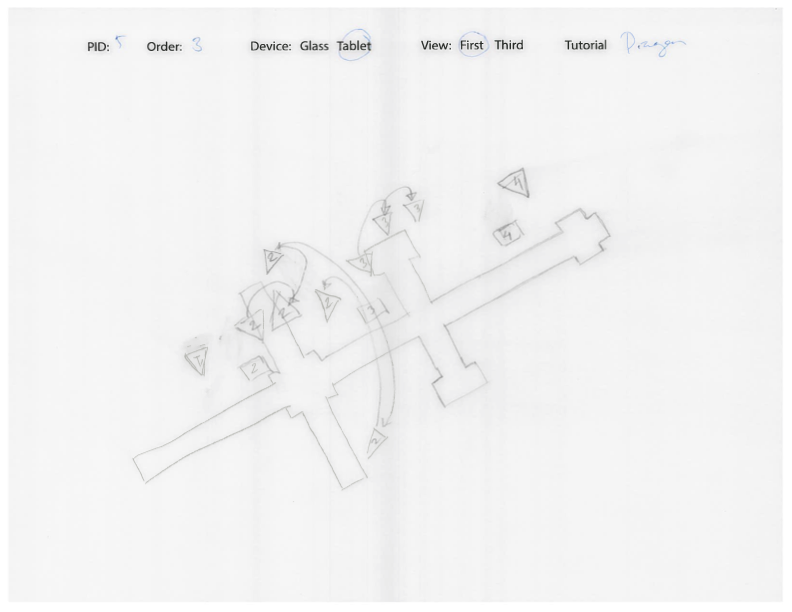} \\
 (a) & (b)  \\
 \includegraphics[width=2.5in]{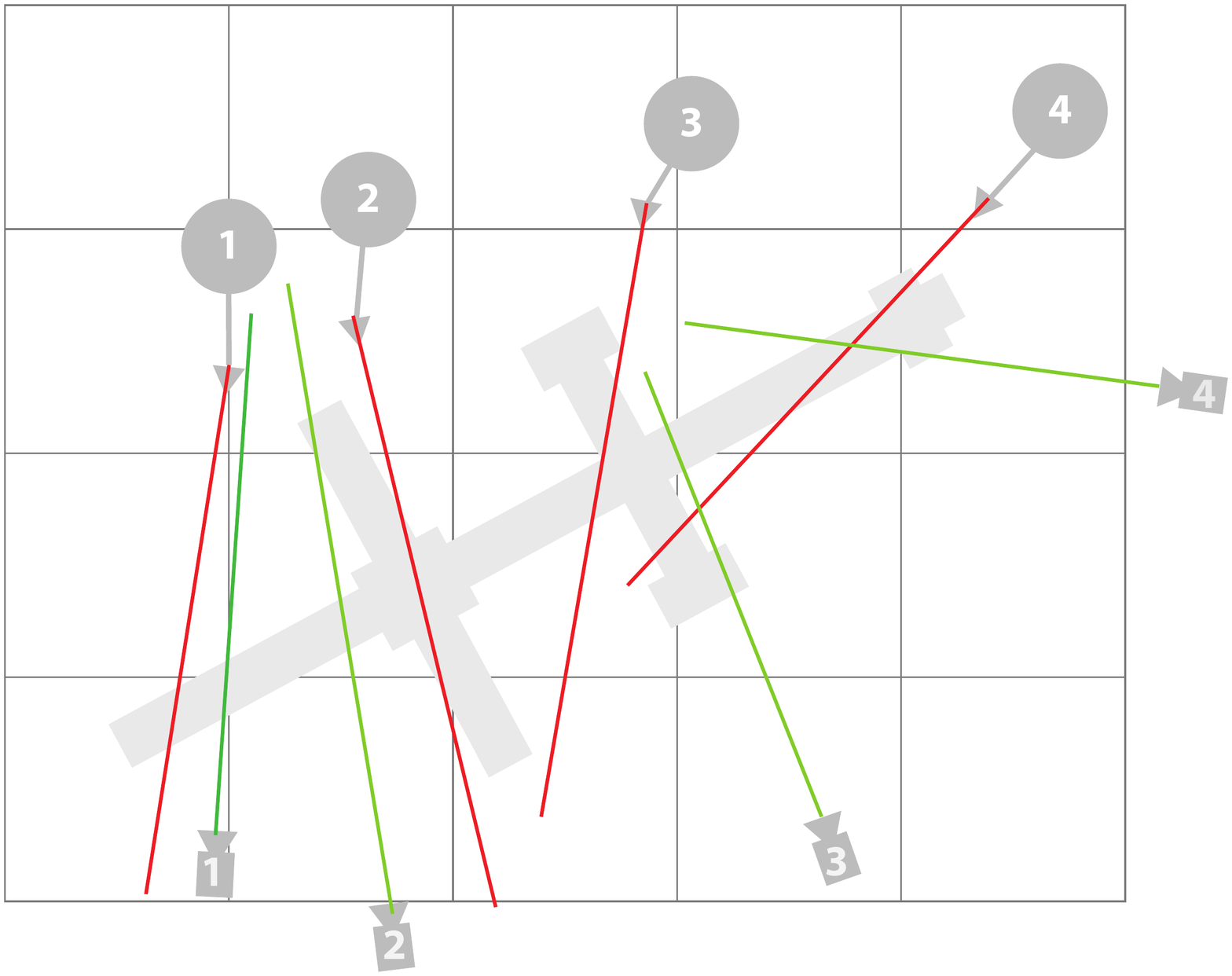} &
 \includegraphics[width=2.5in]{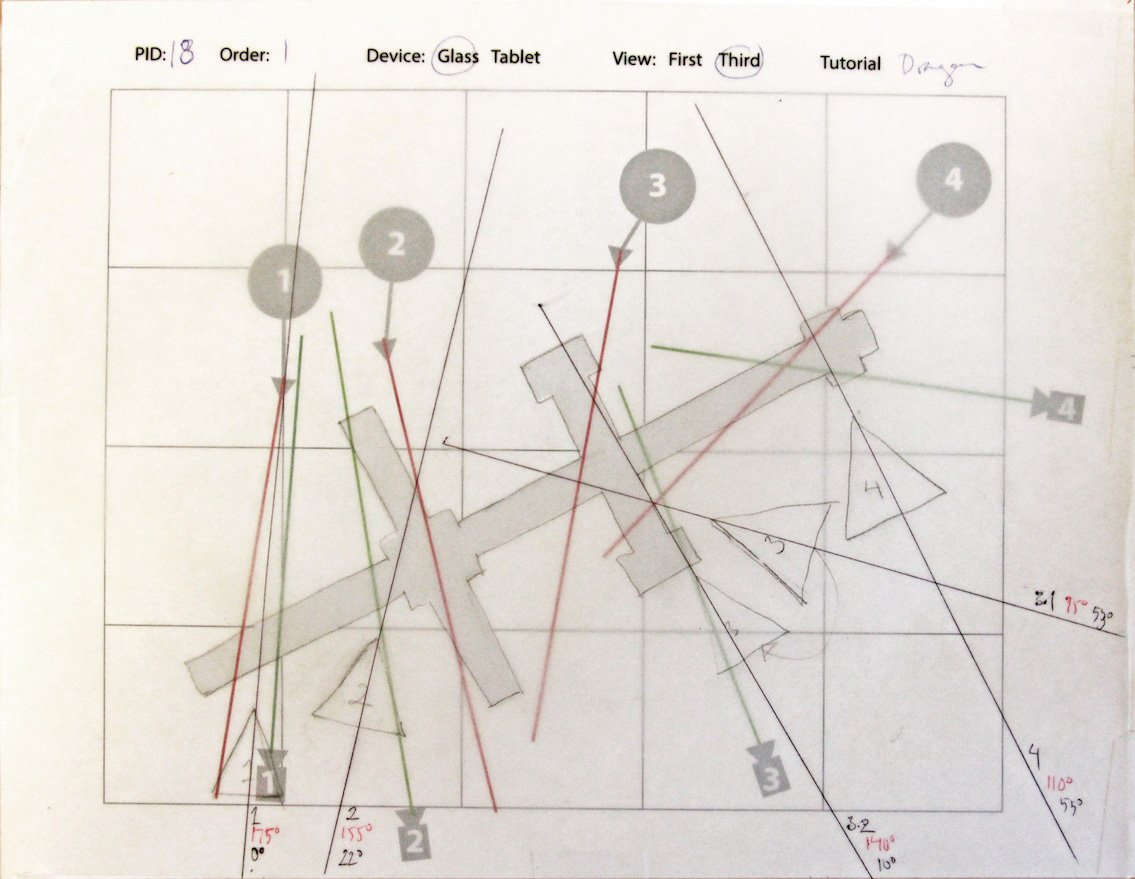} \\
  (c) & (d) \\
 \end{tabular}
 \end{center}
 \caption{Procedure of observing and analyzing participant's position in relation to model (a) Model sketch for reference. Model is divided into segments, one for each step. (b) Observation of a participant's position in relation to model. Triangle indicate participant and rectangle tablet position. The number indicate step. (c) Key with reference positions of tutor and camera on tripod with center view port marked (tutor = red, camera on tripod = green). (d) Reference lines for participants' center view port (in search for a better word)}
 \label{fig:obsProto}
 \end{figure}

Participants' position in relation to the Lego model was recorded through the building process. During recording, the tutors' and cameras' position in relation to the model was recorded, confirmed using the recorded materials and summarized as a key (Figure \ref{fig:obsProto} c). For the observation, a model on a grid was prepared (Figure \ref{fig:obsProto} a). The model was divided into segments corresponding to the steps in the tutorial. The grid was used as a reference coordinate system. The size of one square in the grid corresponded to the size of a carpet square in the room used for recording the tutorials as well as for the study sessions. To recording the participants' building position, a semi-transparent vellum paper was overlaid on the model grid and participants' positions were marked with a triangle there the point of the triangle pointed in the same direction as participants' head. Only position where participants remained for a significant amount of time and where they were building on the model were recorded. A quick turn of the head to locate and pick up a Lego brick for instance was not recorded. Figure \ref{fig:obsProto} (b) shows a completed observation protocol from one of the participants. The triangles are marked with numbers corresponding to the steps in the tutorial. Multiple triangles with the same number indicate multiple positions during that step. Arrows connecting the triangles indicate order of the building positions within a step.

To measure alignment with first- or third-person perspective, the angle between the recording position and the participants' position were measured. The observation protocol was positioned on top of the key and lines were drown with a 90 degree angle against the base of the triangle and through its tip. The angle was measured between the first- and third-person perspective lines and the participant's line and recorded. Figure \ref{fig:obsProto} (d) shows an example of a participants' observation protocol with reference lines added on top of the key.

For the analysis, we used a repeated-measurements ANOVA with the independent variables perspective and access device if no other statistical tests was mentioned. Since no interaction effects were found on any of our dependent variables, we only report data for main effects.
 
\subsection{Results and discussion}

\subsubsection{Performance}
Performance was measured using task completion time and the number of incorrectly placed pieces after the model was completed. We also observed how many corrections or alterations participants made during the building process and at which step the participants made errors and corrections. 

\begin{table}[thb]
\tbl{Overall performance when using tablet and Glass as access device and recording device. Standard Deviations appear in parenthesis.}{
\begin{tabular}{c|c||ccc}
Condition & Levels & Dur. (min) & Errors & Alterations \\ \hline \hline
All &   & 14.0 (4.26) & 2.4 (2.39) & 6.4 (6.80)  \\ \hline 
\multirow{2}{*}{Access}	& Glass  &	15.0 (4.26) & 2.5 (5.05) & 6.4 (7.23) \\ 
						& Tablet &	13.0 (4.11) & 0.9 (2.16) & 6.4 (6.45) \\ \hline
\multirow{2}{*}{Recording}	& First-person &	13.9 (4.03) & 1.4 (3.18) & 7.4 (7.50) \\ 
						    & Third-person &	14.1 (4.56) & 2.1 (4.61) & 5.4 (5.96) \\ 
\end{tabular}}
\label{tab:accessperf}
\end{table}

On average, when the participants used Glass, they spent 15 min (SD=4.26) building the model described in the tutorials (Table \ref{tab:accessperf}). This was on average two minutes longer than when the participants used the tablet. However, this difference was not statistically significant ($F(1, 15)=3.252, ns.$). 

During the building process, we counted the number of alterations the participants made to their model. To count as an alteration, the participants needed to undo parts of model, for instance when they discovered an error. Note that the outcome of an alteration does not need to result in a corrected model. The participants made about the same number of alterations independent of access device and recording view (Table \ref{tab:accessperf}). In contrast, errors were counted after the participants finished building the model. An error was for instance when a Lego block as misplaced or a different Lego block was used. We found that the participants made more errors when accessing the tutorials using Glass (M=2.5, SD=5.05) than when using tablet (M=0.9, SD=2.16). This difference was borderline significant ($F(1, 15)=4.401, p=0.0532$). These results show that we found no performance differences between access devices and recording perspective, except for a higher error-rate when accessing tutorials using Glass.

One reason for a higher error-rate with Glass could be that the access devices demanded different degree of visual attention. We observed that participants primarily used two strategies while accessing tutorials and building the models. Either they first viewed the tutorial and then built the model, or they built in sync with tutorial. None of participants used one of these two strategies during the whole building process. However, the participants claimed that they more frequently built in sync with the video when using Glass than when using the tablet, for instance Participant 16 said ``Regarding tablet, I have to stop [building] the Lego and watch the video, but for the glass, sometimes I can just watch the video while I am doing the Lego.''

Building in sync with the video on the tablet requires the participant to turn their head away from the tablet 
and rely on the audio from the tutorials to guide them. It is possible that when the participants used Glass to access the tutorial, they did not pay as careful attention to their building process as when they were using the tablet. We found evidence of this in our observation of the sessions, where we say participants fumbling to attach one piece to another while looking at Glass' display, and in the participants comments:

\begin{quote}
I like being able to not watch/ignore some bits by not looking at the tablet. It's harder to ignore Glass (Participant 15)
\end{quote} 

\begin{quote}
[It is] hard to pick the right piece and focus on the video. I often got pieces wrong, it's better to watch to see what to do and then build. (Participants 19)
\end{quote}

Another curious result we found was that three participants, all using Glass, built their entire model in a mirror image compared to model in the tutorial without noticing their error. When using a tablet, models were also built in mirror image, however, in all those instances, the participants discovered their mistake and corrected it. Two participants of 16 explicitly pointed out the larger image in the tablet made it easier to compare the tutorial model with their model. Participant 5 said that ``solving the tutorial includes performing two tasks: 1) compare the tutorial view with my view, 2) navigating between the different steps.'' Having a larger image for comparison could make it easier to see discrepancies.

We had expected that Glass would allow participants to glance at the video only as needed and devote their visual attention to model to a greater extent than the tablet.
It seems possible that the presence of moving images in the participants' peripheral view distracted their attention producing in the opposite result. The tablet could be, and was often, placed outside the participant's view.



\subsubsection{Aligning to recording view}
We observed how people aligned and positioned themselves to the model, the tutor, and the recording view while working on the model and recorded each unique building position. When the tutorial was recorded in first-person view, in 50\% of the building positions the participants choose to be within a 45 degree angle from the recording position. When the tutorial was recorded in a third-person view, 42\% of the participants' building position was within 45 degree of the recording view and 20\% was within 45 degree of the tutorial instructors' view. In the rest of the recorded building positions (50\% in first-person view and 38\% in third-person view), the participants situated themselves in a position that was neither aligned to the instructor nor the camera view. Instead they appeared to choose a building position that was close, or conveniently located, to the pile of Lego blocks, or at location from where they could reach the model and limit their required movement.
 Participant 9 explained that he selected a position that allowed him ``to move the least, especially with the tablet... I didn't mind mental rotation. I was optimizing for building fast without needing to pause the tutorial.''  Other participants were very conscious about the recording view and preferred to be aligned with it, as Participant 18 ``I turn maps around, I positioned myself according to the view of the video'' and Participant 19 expressed ``I tried to align with the view in the video so it would be easier to build.''

In addition, we found that the participants aligned themselves significantly closer to a first-person view ($M=55 degrees, SD= 50.0$) than to a third-person view ($M=75 degree, SD=60.1; F(1, 15)=6.045, p<0.05$). The participants' comments gave few insights as to why they would align closer to a first-person view. We found no effect on access device or interaction between access device and recording view. This means that participants using tablet aligned themselves to an equal degree with the recording view as those using Glass. 

Error and the complexity of the model affected alignment with recording view. We found that participants were significantly less aligned to the recording view in steps where they made errors ($F(1, 15)=5.100, p<0.05$). We found no effect on recording perspective or access device. In addition, we found a significant multiple linear regression with $R^2=0.139$, where a higher average degree of alignment with the recording view resulted in fewer errors and fewer modifications ($F(2, 61)=4.927, p<0.05$). Participants also claimed in their comments that when they were more closely aligned, they made fewer errors, for instance Participant 19 said ``when I follow the view I make less errors'' and Participant 4 stated ``If I had difficulty with the orientation of the Lego model, I adjusted my position to match the video. If I didn't have any problem, I didn't care about my position.''

From these results, we can see that alignment with recording view is not necessarily consistent over the a session. Several aspects influence the choices the participants make when selecting a building position. Alignment with recording view is only one. Others are closeness to work space and material to optimize for speed, complexity of the model, and a match with verbal instructions, as Participant 15 explains:

\begin{quote}
I don't have a problem with a shift of view up to 90 degrees, but when it's 180 degrees it bothered me. I wanted the words to match `here to the left' etc with what I saw. I guess I could have ignored the words, but I didn't. (Participant 15)
\end{quote}

We did not find any statistical evidence that participants aligned more closely to the recording view when using one access device over the other. The participant's experience of devices gave diverse preferences and motivations. For instance, Participant 10 provides an interesting perspective on how the access device affected how he aligned his view with the tutorial's view: ``I didn't really think about the orientation of the tutorial, but it's easy with the tablet since I can just rotate it. I found it kind of disturbing with Glass because you had to rotate yourself.'' However, more participants experienced that aligning to the recording view with Glass was more natural:

\begin{quote}
[The] surprise was that I didn't feel that I had to align my view when I had Glass. (Participant 14)
\end{quote}

\begin{quote}
It just wasn't obvious to me how to align when using the tablet (Participant 16)
\end{quote}




The participants expressed very mixed preferences for first- or third-person perspective. Although Figure ~\ref{fig:preference_access} shows that most participants (8) preferred tablet + third-person perspective, two of these participants had not noticed any difference in view while working with the tutorials and two other participants wished to have media shown from multiple perspectives. The most common reason for preferring a first-person perspective was that the participants did not feel a need for compensate for mismatch to the tutor's view (4/5 preferring first-person perspective). The participants preferring a third person view 
reported that third-person perspective gave a better overview and sense of orientation of the model (5/11 preferring third-person perspective). This is particularly interesting since the camera on Glass is designed with a wide angle to match the field of view of the person wearing Glass. Another complaint was that video from Glass was not as stable as the camera on tripod video (2/11). The two participants indicating this was an issue discussed why a stable image is important:

\begin{quote}
Having a fixed viewing angle makes it easier to look and compare with the physical model. Recording with Glass makes the view continuously changing and this makes it very difficult to compare with one glance. (Participant 5 on preferring third-person view)
\end{quote}

\begin{quote}
I found I was slightly behind the person in the tutorial in each step so at the point I am about to place a block they are already looking in another direction grabbing the next piece. A third person view lets me see where they put the last block for a longer period of time. (Participant 1 on preferring third-person view)
\end{quote}

Note that the participants preferring a first-person view and participants preferring a third person view have different focus when accessing the content. Participants preferring a first-person view focus on a match between the view of the tutor and themselves, while participants preferring a third-person view focus on the context shown in the video, orientation of model, a image stability etc. 

\subsubsection{User experience of the access device}
Next we examine participants' experience and preferences for the two access devices. Table \ref{tab:accessratings} shows the participants average rating on confidence building the Lego models, confidence in access device, the ease of following direction, and ease of navigation. Overall, the participants had a more positive experience when using the tablet compared to Glass. When using the tablet, the participants rated their confidence in building the Lego models as significantly higher than when using Glass (Tablet: M=5.8, M=1.13; Glass: M=5.2, SD= 1.67; $F(1, 15)=6.356, p<0.05$) as well as their confidence in access device (Tablet: M=5.8, SD=1.02; Glass: M=4.7, SD=1.08; $F(1, 15)=52.826, p<0.001$). 
Also, participants rated tutorials easier to follow when using tablet (M=5.4, SD=1.65) compared to Glass (M=4.3, SD=1.87; $F(1, 15)=4.694, p<0.05$), and significantly easier to navigate (Glass: M=4.7, SD=1.08; Tablet: M=5.8, SD= 1.02; $F(1, 15)=7.918, p<0.05$). We found no differences in any of these user experience ratings between the two recording views.  

\begin{table}[t]
\tbl{Average user ratings on confidence building Lego model, confidence in access device, ease of following direction and ease to navigate. Standard deviation within parenthesis, * signifies $p < 0.05$ and *** $p < 0.001$.}{%
\begin{tabular}{c|c||c|c|c|c}

\multirow{2}{*}{Condition} & \multirow{2}{*}{Levels} & Confidence & Confidence & Ease of following  & Ease of  \\
 & & building & access device & directions & navigation \\
\hline \hline
All & & 5.5 (1.18) & 5.2 (1.19) & 5.2 (1.44) & 4.8 (1.83) \\ \hline
\multirow{2}{*}{Access} & Glass & 5.2 (1.17) * & 4.7 (1.08) *** & 4.3 (1.87) * & 4.3 (1.87) * \\ 
 & Tablet & 5.7 (1.13) * & 5.8 (1.02) *** & 5.4 (1.65) * & 5.3 (1.66) * \\ \hline
\multirow{2}{*}{Recording} &  First-person & 5.5 (1.05) & 5.3 (1.10) & 4.7 (1.81) & 4.8 (1.81) \\ 
 & Third-person & 5.4 (1.32) & 5.2 (1.29) & 4.3 (1.81) & 4.9 (1.87)
\end{tabular}}
\label{tab:accessratings}
\end{table}


\begin{figure}[thb]
 \centering
 \begin{tabular}{c} 
  \includegraphics[width=3.5 in]{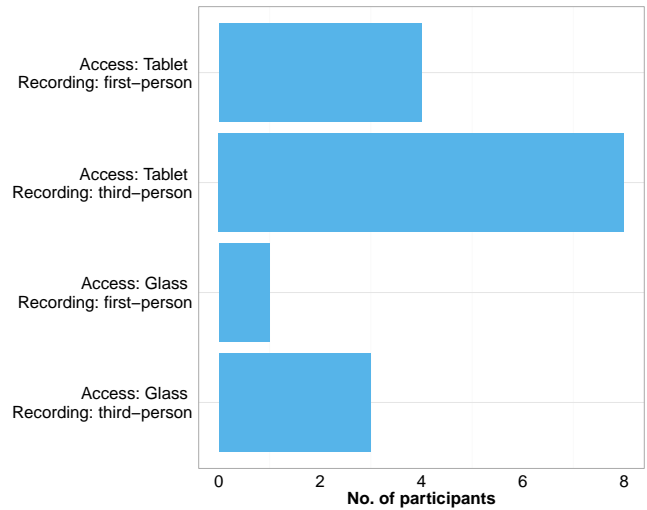} \\
 \end{tabular}
 \caption{Number of participants preferring combinations of tablet or Glass for access and first or third person view for recording.}
 \label{fig:preference_access}
 \end{figure}

When asked about a particular combination of recording and access device they preferred, most participants (8) preferred to view a tutorial recorded with a third person view on a tablet (see Figure \ref{fig:preference_access}), although two of these had not noticed any difference in perspective and had no strong preferences. Four participants preferred to have a first person view shown on a tablet. The reasons participants gave for preferring tablet over Google Glass was that Glass was more difficult to interact with than the tablet (5/11 preferring tablet) and that photos and video were larger on a tablet (2/11) which facilitates comparing the model they are building with the tutorial. Thirteen of sixteen participants mentioned that Glass was unresponsive or overly sensitive. This is partly because Glass at the time got unresponsive when hot. In addition, Glass is uncomfortable to use for watching longer videos (3/16). Some people have trouble switching focus between the near eye display and the more distant Lego model. Glass had some positive aspects. A couple of participants enjoyed the hands-free aspect of Glass (2/16), and the convenience of carrying the display close to the eyes (3/16.) Three participants acknowledge that they grew more confident using Glass the second time, and it is possible that with longer exposure to Glass the participants may have figured out how to trigger the intended action.

The tablet player had some negative aspects as well. Four of sixteen participants had some issues interacting with the tablet. In particular the fast forward and fast reverse was hard to perform. The gesture for moving to the next step and fast reverse is very similar, the only difference is the use of one finger versus two fingers. This seem to confuse the participants both when interacting with tablet and Glass. In addition, some participants had trouble understanding why moving to the next step (swipe left on tablet) and moving forward within the video (two finger swipe right) included gestures in opposite directions.

Compared to a tablet, Google Glass intuitively has some advantages. The display is wearable, leaving both hands free to interact with the object of interest. However, in our study the participants overwhelmingly preferred to use a tablet when using a tutorial to recreate a model. The reasons the participants gave were mostly centered on the hardware. Many participants experienced interaction issues with Google Glass. These stem from two causes, an oversensitive touch display and an insensitive touch sensor when Glass got too hot. The oversensitive touch display was particular bothersome when a tap was interpreted as a swipe, causing the participant to move to the next step instead of just pausing the video. This in turn meant that the participant needed multiple steps to get back to the location in the video they wished to pause (move back, fast forward etc). Participant 1 exemplifies what many  experienced:

\begin{quote}
I fell behind on one of the steps and tried to move the video back but instead jumped back about 3 steps, having to find the correct step and then start that step from the beginning (Participant 1 on his experience using Glass)
\end{quote}

\subsubsection{Conclusions and Discussion}
In this study, we explored how well online tutorials can be accessed using two different kinds of devices, Google Glass and a Nexus 7 tablet, running almost identical software in mobile setting.  In Table \ref{tab:accessprocon} we have  summarized both positive and negative aspects of the two access devices. Our users generally found the tablet easier to use than Glass, mainly due to Glass hardware issues. These hardware issues are likely to be resolved in the future, and user preference may change with better hardware. Our participants’ indicated that Glass could have worked out well for accessing tutorials if these hardware issues were solved:  

\begin{quote}
The Glass Tutorial would have work out fine if not for difficulty in tapping to stop the video. (Participant 8)
\end{quote}

\begin{table}[t]
\tbl{Pros and cons for using a tablet or Google Glass to access tutorials.}{
\begin{tabular}{c||p{5.4cm}|p{5.4cm}}
Access device & Pros & Cons \\ \hline \hline
 Tablet & Larger size media &  May need to turn attention away from work space\\ 
 		& Can be turned to align with view	& Needs to be placed or hold in hand \\
        & Easy to navigate & Restricted building position \\
        & Can be ignored & \\ \hline
Glass & Easily aligned with recording perspective & Hardware issues \\ 
      & No need to turn attention away & Eyes gets tired \\
      & Unlimited building positions  & Video attracts too much attention \\
      & Hands-free & \\
      & Information quickly accessible with a glance & 
\end{tabular}}
\label{tab:accessprocon}
\end{table}

Although future versions of Glass hardware have the potential to change the user experience, our study indicates that accessing online tutorials using a head-mounted display faces other design challenges beyond hardware improvements.  Our study highlights in particular two challenges: limited interaction space and demand on visual attention. 

Our approach for designing the interaction was to map steps in the tutorials to cards that the user could progress by a single finger swipe. This generally worked well; in tutorials users expect distinct steps of instruction. In our system, each step can include a video clip. The users could pause/start video with a tap on Glass or Tablet, or do a two finger swipe to seek 10 sec forward or backwards. A common request was to have larger control over the seek function than in discrete 10 seconds steps. However, the size of the touch sensitive area on Glass is limited, which makes it hard to control a continuous seek in particular for slightly longer video clips. Some participants also confused the single and double finger swipe. The scope of available gestures to control content via a small touch sensor is limited, and this is an issue for creating easy to use gestures. 

We found that our participants made more errors that were not corrected when using Glass compared to using the tablet. Our interpretation, supported by participants' comments, is that video in the users' peripheral vision when using Glass demanded more attention than a video displayed on a tablet out of view. This issue may be possible to address with changes in the design of the tutorial viewer or to the tutorials themselves. Since the problem seems to arise from video, there is a range of possible solutions. A simple option is to encourage tutorial authors to use photos over videos, or to automatically break up the video clips in smaller segments. When a break point is reached, the tutorial would pause until the user gives a go ahead command. A more  involved approach is to extract frames from video clips to show as a narrated slideshow instead of the video. This option would allow the authors to create their tutorial as they wish, but the tutorial is automatically optimized for the access device.

Some of these proposed solutions to lower the visual demands of Glass could also be used to handle the quality issues with video clips recorded with Glass. Tutorial authors greatly preferred to use Glass for a seamless capture of tutorials. However, video captured with Glass is shaky and may have rapid framing shifts when the authors look away to pick up a tool or building material. These quality issues were disturbing to our tutorial viewers and possibly increased the visual distraction.  A framing shift may signify natural break point in a video clip and can be fairly easy to detect with video analysis methods. By pausing the video clip at, or right before, a framing shift, tutorial users can focus on finishing their current task and when they are ready to focus on the video, they can resume playback. One alternative solution is to replace parts of the video with poor quality with one or a set of higher quality frames from right before or during the problematic segment. If the video clip includes narration during a low quality video segment, this method might be preferred over the break point method. 

We believe that before discarding Glass, or other head-mounted devices, for tutorial access, carefully designed tutorial options should be further explored. Yet our results show that the higher demand on visual resources is a challenge. Users reported that they more often built in sync with the tutorial video while using Glass than the tablet. A quick glance is good for confirmation that they are on the right path, but constantly checking the video rather than what they build is not desirable.

Our investigation of how the participants oriented themselves in relation to the recording view and the model they were building showed that with a closer alignment to recording perspective, fewer errors were made. We also found that our participants were more closely aligned to a first-person perspective compared to a third-person perspective. In a first-person perspective, action and language match. Clearly there are benefits to moving closer to the perspective of the tutor. However, participants preferred recordings in third-person perspective. The main objection to the first-person perspective was that the camera image was less stable and had fast framing shifts relative to third-person camera. A few participants pointed out that getting views from multiple perspectives would be helpful. Again our results point to the fact that improved media capture or better selection of media to present in the tutorials is critical for online tutorials.

\section{Conclusion and Future Work}

Creating a tutorial requires capturing descriptive and procedural content and rendering it so that end users can comprehend and recreate the steps taken. In our work, we have seen that seamless capture of tutorial content allows authors to focus on creating high quality content. Studies we ran showed that head-mounted capture is easy to use and configure. Authors also appreciated having a first person view of the content. Our users preferred to capture tutorial content with our Google Glass ShowHow application over a camera on a tripod, even when the camera on tripod included the same convenient features such as automatic content upload. The Glass application was seen as particularly useful for obtaining content from different angles and adding natural narration. Head-mounted capture was preferred for large objects where authors needed to move around the objects to show it from different angles. When handling smaller objects on a desktop, the camera on tripod was seen as equally useful as head-mounted capture. 

A close integration between capture and authoring tools streamlines the tutorial creation process. 
First and foremost, this requires high quality content.  Furthermore, reinforcing content across different mediums both provides additional context and accommodates different users' preferred learning styles. 
We believe that our suite of web tools allows users to combine a variety of different media to convey expository material.
The authoring tool can accept media from various sources including a heads-up capture application for Google Glass that we built. We performed a user study and determined that users found it more intuitive to build tutorials using a multimedia document approach. Furthermore, the multimedia document approach encourages a much more heterogeneous usage of media. We also found that heads-up capture systems are well suited to capturing unwieldy content for tutorials because of their relative unobtrusiveness. 
In addition, we have seen that the conceptual structure of the authoring tool influences the way media is recorded. With a video-based structure and a head-mounted camera, users recorded longer video clips. 
However, a document-based method prompted users to add more photos and annotations to the tutorial independent of capture method. Tutorials with fewer videos and more annotations and bookmarks tended to have higher quality.  
Thus, we have seen that tutorial representation can influence how authors approach authoring, and in turn may affect both how long it takes to make a tutorial and its perceived quality. 


In contrast to authors' strong preference for head-mounted capture, tutorial consumers preferred viewing tutorials with a tablet in a third-person perspective. This change of preference is partly due to hardware issues, but also to our assumptions about how multimedia tutorials are viewed. The reason our authors gave for preferring head-mounted capture is that the display is glanceable. However, when following a tutorial, the user needs to understand what to do and carefully follow directions given. When working on a complex task, it may not be possible to compare reality to a photo or video with a glance. In addition, when the video is placed close to the eye in the peripheral vision, its motion attracts attention and is hard to ignore even when it is not beneficial to view it. A light-weight tablet shows media in a larger format than a head-mounted display and can be placed outside the users' peripheral vision with ease. Yet, head-mounted displays are attractive devices for tutorial consumption in many circumstances. In order to work well, design of the tutorials and the tutorial access application needs a more intricate approach than was supported in our initial design. 
These findings suggest that future work should concentrate on new designs for head-mounted media playback that limit distraction by sensing users' attention and gaze.

Our participants also expressed a preference for a third-person view in the tutorial rather than the first-person view recorded with Glass. In our analysis, on the other hand, we found some benefits with a first-person perspective. People are less likely to make mistakes if they are more closely aligned to the recording view, and people align more closely to a first-person view. However, a first-person view recorded with head-mounted capture suffers from some quality issues that our participants found disturbing, such as rapid shifts in framing. By more carefully taking into account the artifacts of head-mounted capture in the design of authoring and access tools, we may counteract these quality issues.

\bibliographystyle{ACM-Reference-Format-Journals}
\bibliography{tochi}

\newpage 




\end{document}